\newif\ifsupp
\supptrue 

\documentclass[aps,prl,superscriptaddress,twocolumn,10pt,floatfix,nofootinbib]{revtex4-2}

\input{glyphtounicode}
\usepackage[T1]{fontenc}
\usepackage[utf8]{inputenc}
\usepackage[english]{babel}
\usepackage{bm}
\usepackage{amsmath}  
\usepackage{amsfonts} 
\usepackage{amssymb}
\usepackage{graphicx} 
\usepackage{hyperref}
\hypersetup{pdftex,bookmarks=true,bookmarksopen,bookmarksnumbered,
                colorlinks,
                linkcolor=blue,
                citecolor=blue}
\graphicspath{ {./} }
\usepackage{natbib}
\setcitestyle{super}
\usepackage{array}
\usepackage[table]{xcolor}
\usepackage{xcolor}
\usepackage{xr}
\usepackage{braket}
\usepackage{booktabs}
\usepackage{multirow}
\usepackage{dsfont} 
\usepackage{caption}

\usepackage{tabularx}
\usepackage{siunitx}
\usepackage{textcomp}
\usepackage{subcaption}
\usepackage{bigstrut}
\usepackage{svg}
\usepackage{upgreek}

\usepackage{newfloat}

\DeclareFloatingEnvironment[name={\textbf{Supplementary Figure}}]{suppfigure}

\DeclareCaptionLabelSeparator{line}{\hspace{2pt}\bf{|}\hspace{2pt}}

\begin{document}

\captionsetup[table]{name={\bf{Table}},labelsep=period,justification=raggedright,font=small,singlelinecheck=false}
\captionsetup[figure]{name={\bf{Figure}},labelsep=line,justification=raggedright,font=small,singlelinecheck=false}


\renewcommand{\equationautorefname}{Eq.}
\renewcommand{\figureautorefname}{Fig.}
\renewcommand*{\sectionautorefname}{Sec.}

\title{Scalable quantum current source on commercial CMOS process technology}

\author{Ajit~Dash}
\email{ajit.dash@unsw.edu.au}
\affiliation{School of Electrical Engineering and Telecommunications, University of New South Wales, Sydney, NSW 2052, Australia}
\author{Suyash~Pati~Tripathi}
\affiliation{Edward S. Rogers Snr. Department of Electrical and Computer Engineering, University of Toronto, Toronto, ON M5S 3G4, Canada}
\author{Dimitrios~Georgakopoulos}
\affiliation{National Measurement Institute, Australia, West Lindfield, Sydney, NSW 2070, Australia}
\author{MengKe~Feng}
\affiliation{School of Electrical Engineering and Telecommunications, University of New South Wales, Sydney, NSW 2052, Australia}
\affiliation{Diraq Pty. Ltd., Sydney, NSW 2052, Australia}

\author{Steve~Yianni}
\affiliation{School of Electrical Engineering and Telecommunications, University of New South Wales, Sydney, NSW 2052, Australia}
\affiliation{Diraq Pty. Ltd., Sydney, NSW 2052, Australia}
\affiliation{Current address: CEA- Interdisciplinary Research Institute of Grenoble, Grenoble 38054, France}

\author{Ensar~Vahapoglu}
\affiliation{School of Electrical Engineering and Telecommunications, University of New South Wales, Sydney, NSW 2052, Australia}
\affiliation{Diraq Pty. Ltd., Sydney, NSW 2052, Australia}
\author{Md~Mamunur~Rahman}
\affiliation{School of Electrical Engineering and Telecommunications, University of New South Wales, Sydney, NSW 2052, Australia}
\author{Shai~Bonen}
\affiliation{Edward S. Rogers Snr. Department of Electrical and Computer Engineering, University of Toronto, Toronto, ON M5S 3G4, Canada}
\author{Owen~Brace}
\affiliation{National Measurement Institute, Australia, West Lindfield, Sydney, NSW 2070, Australia}
\author{Jonathan~Y.~Huang}
\affiliation{School of Physics, University of New South Wales, Sydney, NSW 2052, Australia}
\author{Wee~Han~Lim}
\affiliation{School of Electrical Engineering and Telecommunications, University of New South Wales, Sydney, NSW 2052, Australia}
\affiliation{Diraq Pty. Ltd., Sydney, NSW 2052, Australia}
\author{Kok~Wai~Chan}
\affiliation{School of Electrical Engineering and Telecommunications, University of New South Wales, Sydney, NSW 2052, Australia}
\affiliation{Diraq Pty. Ltd., Sydney, NSW 2052, Australia}
\author{Will~Gilbert}

\affiliation{Diraq Pty. Ltd., Sydney, NSW 2052, Australia}
\author{Arne~Laucht}
\affiliation{School of Electrical Engineering and Telecommunications, University of New South Wales, Sydney, NSW 2052, Australia}
\affiliation{Diraq Pty. Ltd., Sydney, NSW 2052, Australia}
\author{Andrea~Morello}
\affiliation{School of Electrical Engineering and Telecommunications, University of New South Wales, Sydney, NSW 2052, Australia}
\author{Andre~Saraiva}

\affiliation{Diraq Pty. Ltd., Sydney, NSW 2052, Australia}
\author{Christopher C. Escott}

\affiliation{Diraq Pty. Ltd., Sydney, NSW 2052, Australia}
\author{Sorin~P.~Voinigescu}
\affiliation{Edward S. Rogers Snr. Department of Electrical and Computer Engineering, University of Toronto, Toronto, ON M5S 3G4, Canada}
\author{Andrew~S.~Dzurak}
\affiliation{School of Electrical Engineering and Telecommunications, University of New South Wales, Sydney, NSW 2052, Australia}
\affiliation{Diraq Pty. Ltd., Sydney, NSW 2052, Australia}
\author{Tuomo~Tanttu} 
\email{t.tanttu@unsw.edu.au}
\affiliation{School of Electrical Engineering and Telecommunications, University of New South Wales, Sydney, NSW 2052, Australia}
\affiliation{Diraq Pty. Ltd., Sydney, NSW 2052, Australia}

\date{\today}
\noindent\rule{0pt}{0pt}%
\vspace{0pt}

\begin{abstract}

\textbf{Many quantum technologies require a precise electrical current standard that can only be achieved with expensive cryogenics, or through the secondary standards, such as resistance or voltage. Silicon-based charge pumps could provide such a standard in an inherently scalable way, through their compatibility with complementary metal-oxide-semiconductor (CMOS) fabrication methods. However, coherent quantized charge transfer has so far been demonstrated only in nanoscale devices that are custom-fabricated in academic cleanrooms or research technology foundries. Here, we show that a CMOS device manufactured with commercial 22-nm process node can be used to define a quantum current standard in the International System of Units (SI). We measure an accuracy of $\bm{(1.2 \pm 0.1) \times 10^{-3}}$~A/A at 50~MHz with reference to SI voltage and resistance standards in a pumped helium system. We then propose a practical monolithic CMOS chip that incorporates one million parallel connected charge pumps along with on-chip control electronics. This chip could be operated as a table-top primary standard that can be easily integrated with CMOS electronics, generating quantum currents of up to microampere levels.}


\end{abstract}

\maketitle

\begin{figure*}[hbt!]
    \includegraphics[width=1\textwidth,angle = 0]{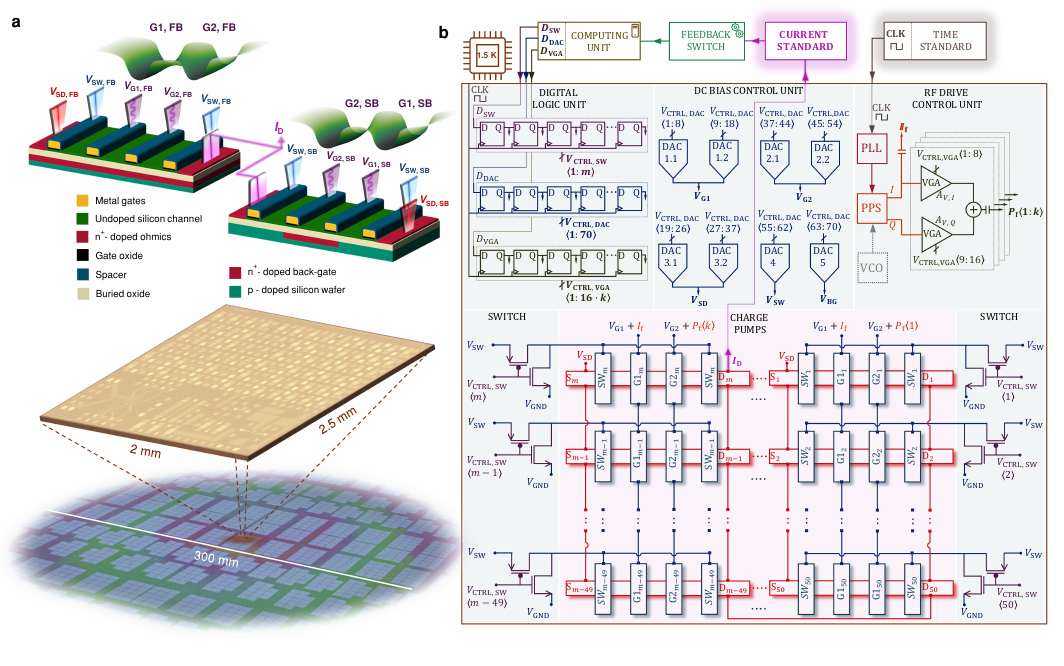}
    \caption{\textbf{Device and proposed monolithic IC architecture}. \textbf{a}, Schematic of the full back-gate (FB) and selective back-gate (SB) devices fabricated using the 22-nm fully-depleted silicon-on-insulator (FDSOI) process technology, the 2~mm $\times$ 2.5~mm integrated circuit (IC) die, and the 300~mm wafer. The FB and SB devices are parallel-connected, sharing a common drain terminal, internally sorted within the IC die. Both the devices consists of a double quantum dot formed under dot gates (G1 and G2). \textbf{b}, Proposed system architecture of the scalable quantum current standard, which is traceable to the international system of units time standard. The IC features monolithic integration of one million parallelized charge pumps with on-chip control electronics, compatible with commercial FDSOI process technology.}
    \label{fig:fig_1}
\end{figure*}

\begin{figure*}[hbt!]
    \includegraphics[width=1\textwidth,angle = 0]{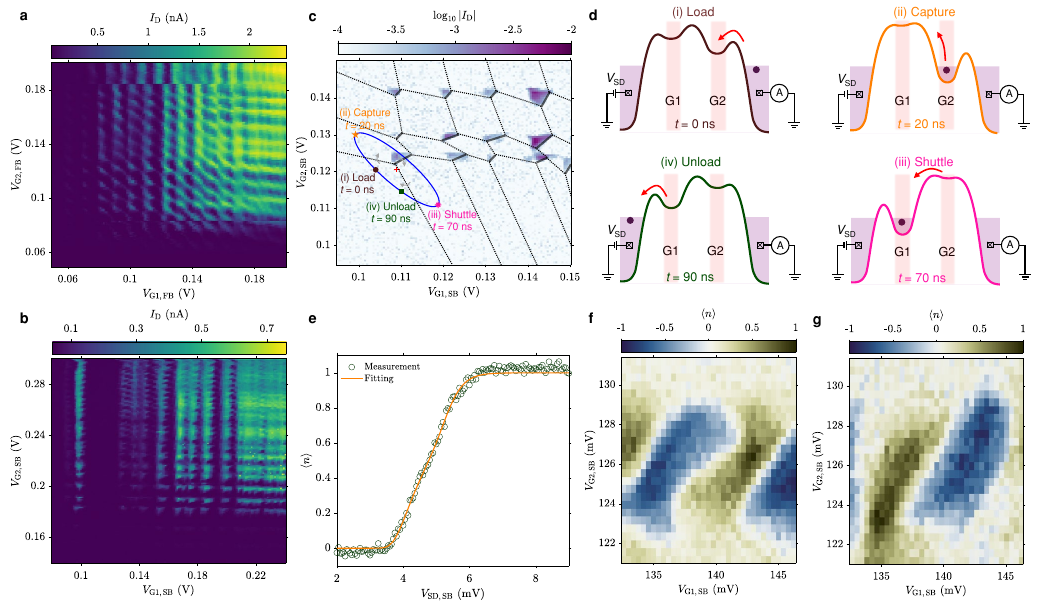}
    \caption{\textbf{Quantized electron pumping (T=1.5 K).} \textbf{a}, Common drain terminal current ($I_{\mathrm{D}}$) as a function of full back-gate (FB) device dc voltages $(V_\mathrm{{G1,\,FB}}$, and $V_\mathrm{{G2,\,FB}})$ and \textbf{b}, selective back-gate (SB) device dc voltages $(V_\mathrm{{G1,\,SB}}$, and $V_\mathrm{{G2,\,SB}})$. The common back-gate terminal voltage $(V_\mathrm{{BG}})$ is supplied with 3.5~V for both FB and SB devices throughout all the measurements. \textbf{c}, $I_{\mathrm{D}}$ as a function of $V_\mathrm{{G1,\,SB}}$, and $V_\mathrm{{G2,\,SB}}$ at SB device source-drain bias, $V_\mathrm{{SD,\,SB}}$=3~mV, superimposed with the Lissajous curve trajectory (blue curve) of two rf waveforms $\widetilde{V}_{\mathrm{G1,\,SB}} \sin\left(2\pi f_{\mathrm{SB}} t + \phi_{\mathrm{G1,\,SB}}\right)$, and $\widetilde{V}_{\mathrm{G2,\,SB}} \sin\left(2\pi f_{\mathrm{SB}} t + \phi_{\mathrm{G2,\,SB}}\right)$. The Lissajous curve is centered at the operation triple point (red plus) $V_\mathrm{{G1,\,SB}}$=109.0~mV, and $V_\mathrm{{G2,\,SB}}$=120.5~mV, $\widetilde{V}_{\mathrm{G1,\,SB}}=\widetilde{V}_{\mathrm{G2,\,SB}}$=10~mV, $f_{\mathrm{SB}}$=10~MHz, $\Delta \phi_\mathrm{SB} = \phi_{\mathrm{G1,\,SB}}-\phi_{\mathrm{G2,\,SB}}=-150^\circ$. The brown circle ($t$=0~ns), orange star ($t$=20~ns), pink asterisk ($t$=70~ns), and green square ($t$=90~ns) indicate the four sequences of single-electron pumping across the double quantum dot, during the first frequency cycle. \textbf{d}, Time-dependent electrostatic simulation of the FDSOI device at t=0~ns, 20~ns, 70~ns, and 90~ns, plotting the potential profile during the four stages of single-electron pumping. \textbf{e}, Average number of pumped electrons $\langle n \rangle$ per 10~MHz frequency cycle of the two phase-shifted rf waveforms (blue curve, in \textbf{c}) as a function of $V_\mathrm{{SD,SB}}$, operated at the operation triple point (red plus, in \textbf{c}). \textbf{f-g}, Bidirectional electron pumping map $\langle n \rangle= \pm1$ for $\Delta \phi_{\mathrm{SB}}=90^\circ$ and $\Delta \phi_{\mathrm{SB}}=0^\circ$.
    }
    \label{fig:fig_2}
\end{figure*}

\begin{figure}[hbt!]

    \centering
    \includegraphics[width=0.47\textwidth,angle = 0]{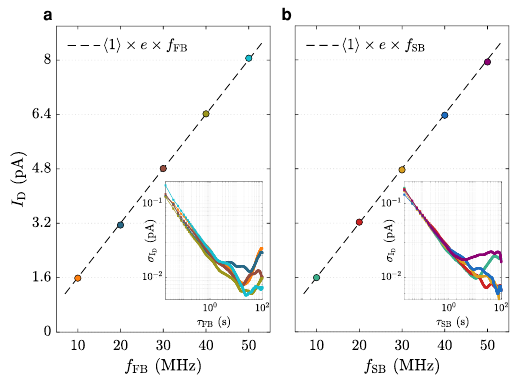}
    \caption{\textbf{Frequency dependence of quantized current (T=1.5 K).}
    Pumped current measured at the common drain terminal ($I_{\mathrm{D}}$), labeled as colored circles, as a function of rf waveform frequency for full back-gate (FB) device in \textbf{a} and, selective back-gate (SB) device in \textbf{b}. Plotted $I_{\mathrm{D}}$ is obtained by averaging 2500 data points at the flattest , each measured with an integration time of 20 ms and a wait time of 100 ms. To compensate for the measurement circuit offset current, correction values of 0.2~pA and 0.05~pA are added to the raw data of the FB and SB devices, respectively. The black dashed line plots the expected current, $\langle 1 \rangle \times e \times f_{\mathrm{FB}}$ in \textbf{a}, and $\langle 1 \rangle \times e \times f_{\mathrm{SB}}$ in \textbf{b}. \textbf{Insets}: Allan deviation of pumped current ($\sigma_{I_{\mathrm{D}}}$) at different rf waveform frequencies $f_{\mathrm{FB}}$ in \textbf{a} and, $f_{\mathrm{SB}}$ in \textbf{b}. The color coding used for the frequency dependent $\sigma_{I_{\mathrm{D}}}$ data curve matches the colored circles representing $I_{\mathrm{D}}$. The Allan deviation curves for both FB and SB devices shows a $\tau_\mathrm{FB/SB}^{-1/2}$ dependence, indicating that white noise is the dominant factor up to integration times of about 1~s.
    }
    \label{fig:fig_3}
\end{figure}

\begin{figure*}[hbt!]

    \centering
    \includegraphics[width=0.85\textwidth,angle = 0]{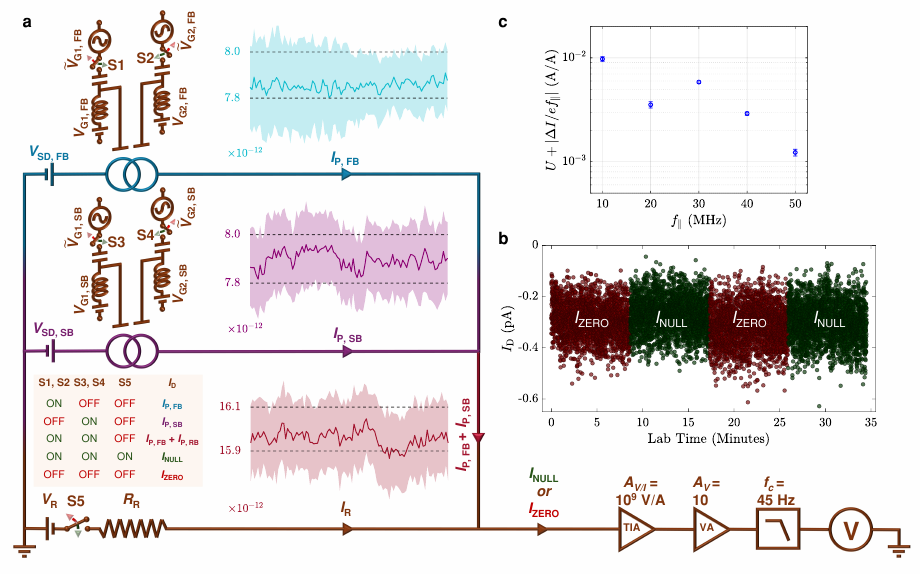}
    \caption{\textbf{Parallel operation and accuracy measurements (T=1.5 K).}
    \textbf{a}, Electrical circuit schematic of the setup used for parallel pumped operation and accuracy measurement of summed quantized current. Full back-gate (FB) and selective back-gate (SB) devices controlled with tuned dc voltages and rf waveforms, function as quantized electron pumps. The common drain terminal architecture sums the current $(I_{\mathrm{D}}=I_{\mathrm{D,\,FB}}+I_{\mathrm{D,\,SB}})$ when both the electron pumps operate simultaneously. The data traces in \textbf{a}, are obtained by averaging 500 data points, with the shaded regions indicating the standard deviations; each measured with an integration time of 20 ms and a wait time of 100 ms, at an rf waveform frequency of 50 MHz. Individual traces show pumped current of approximately 8 pA through the FB device (cyan), 8 pA through the SB device (magenta), and about 16 pA when both devices operate simultaneously (burgundy). A reference output voltage of reverse polarity ($V_{\mathrm{R}}$) is applied to a reference resistor ($R_{\mathrm{R}}$ = 1.00936135 G$\Omega$) to generate a negative reference current $(I_{\mathrm{R}})$ for accuracy measurements. \textbf{b}, Measured null current ($I_{\mathrm{null}}$) and zero current ($I_{\mathrm{zero}}$) over two alternating cycles while simultaneously pumping single-electrons in parallel using both the FB and SB devices with parallel pump frequency $f_{\parallel}$=50~MHz. \textbf{c}, Total relative uncertainty $U+|\Delta I/ef_{\parallel}|$ and its estimated standard deviation $(\sigma_{\Delta I/ef_{\parallel}})$ of the parallel pump current, with respect to $I_{\mathrm{R}}$, as a function of $f_{\parallel}$. Each data point is obtained by averaging 5000 data points, measured with an integration time of 20~ms and a wait time of 100~ms.}
    \label{fig:fig_4}
\end{figure*}

As technologies become ever more precise, measurement standards must evolve \cite{tzalenchuk2022expanding, kuramoto2022new, stock2019revision}. The 2019 revision of the SI was one step in this evolution, fixing the values of fundamental physical constants in a way that was intended to be more accessible than the previous definitions\cite{stock2019revision,BIPM_SI_2019}. For example, one of the seven base units, the ampere, was redefined in terms of the fixed elementary charge $e$, which relates to the SI time standard through frequency $f$ \cite{BIPM_SI_2019, stock2019revision, mohr2018data, schulte2020prospects, dash2024silicon, djordjevic2025primary}.


However, the accurate realization of electrical currents typically requires experiments conducted at temperatures close to absolute zero, making it inaccessible to most scientists and engineers, who instead must rely on measuring current standards with respect to other standard quantities. Precise current metrology is crucial, for example, in varied applications such as space research, nuclear radioactivity monitoring, air quality mapping, and medical instrumentation involving X-rays and gamma rays, including radiation oncology\cite{kaneko2023perspectives,fitzgerald2020next,okazaki2022subfemtoampere}. Accurate generation of this quantum current would also close the quantum metrological triangle\cite{feltin2009determination, scherer2012quantum, devoille2012quantum, scherer2019single, fujiwara2023silicon} — allowing determination of the consistency between independently realized quantum standards of current, voltage, and resistance. Moreover, it would also enable advances in technologies that require traceable calibration of precise electric currents\cite{tzalenchuk2022expanding,kaneko2023perspectives}, such as utility-scale quantum computers.

A silicon-based charge pump offers a way of making this current standard accessible and — importantly — easily integrated with classical electronics. A charge pump is a nanoscale device that transfers $\langle n \rangle$ integer electrons per clock cycle, yielding a current of $I=\langle n \rangle \times e \times f$\cite{BIPM_Ampere_2019}. Silicon-based quantum dots (QDs) offer a promising way to implement such devices because they have high charging energies, allowing them to operate at temperatures of up to a few kelvin\cite{giblin2020realisation, giblin2023precision, rossi2018gigahertz, dash2024silicon}. However, existing charge pumps achieve either high speeds or high fidelity, not both \cite{kataoka2011tunable, ahn2017upper, johnson2024high, akmentinsh2025modeling}. As a result, the maximum quantized current output with metrologically relevant accuracy is limited to the range of hundreds of picoamperes \cite{stein2016robustness, yamahata2016gigahertz, zhao2017thermal}.

The solution is to operate multiple charge pumps in parallel to increase the total output current while maintaining quantization accuracy \cite{kim2022tuning, nakamura2023universality, schoinas2024fast, norimoto2024statistical, yamahata2025scalable}. However, scaling up the number of charge-pumps also increases the complexity of the control infrastructure, particularly the number of required control lines, thus creating a bottleneck in scalability. Fabricating a large number of QD devices on a single chip can also exceed the capabilities of in-house resources. Leveraging the scale and reliability of commercial CMOS foundries offers a viable and scalable path toward overcoming these challenges.


\section*{Commercial quantum dot device}
In this work, we demonstrate quantized electron pumping using two parallel-connected FDSOI devices integrated on the same integrated circuit (IC) die, sharing a common drain terminal. The IC die (Fig.~\ref{fig:fig_1}a) was fabricated using the commercial GlobalFoundries\textsuperscript{\texttrademark} 22FDX\textsuperscript{\textregistered} platform.  Both devices measured in this study consist of an array of four high-$k$ metal gate stack, each with dimensions, $L$ $\times$ $W$ = 28~nm $\times$ 50~nm and an experimental gate pitch, $P$ = 90~nm. Each device has a silicon channel thickness of 6 nm, buried oxide thickness of 20~nm and ESD-protected top-gates via 24~dB attenuators connected to each fan-out bond pad. This facilitates the monolithic integration of QD devices with conventional CMOS circuitry on the same chip\cite{bonen2024investigation,zhao202380}. 

When operated at cryogenic temperatures, these devices naturally exhibit discrete occupation states~\cite{bonen2018cryogenic,bonen2024investigation,tripathi2022characterization,amitonov2024commercial,thomas2025rapid, elbaz2025transport}, as the tunneling resistance of the QD is higher than the resistance of quanta ($\mathrm{R}_{t} >\mathrm{R}_{\mathrm{K}}$) and the charging energy is larger than thermal energy ($\mathrm{E}_{\mathrm{C}} \gg k_{\mathrm{B}}\mathrm{T}$). Double quantum dot (DQD) islands are electrostatically defined within the undoped silicon channel, under the dot gates $(\mathrm{G1 \,and \, G2})$. Gated spacers, together with the ungated regions, define tunnel barriers for the QD islands. These barriers are floating and are tuned by adjusting the neighboring top-gate and the back-gate voltages ~\cite{tripathi2022characterization,amitonov2024commercial}. The n$^{\mathrm{+}}$ doped silicon layer below the buried oxide features the back-gate, as a secondary gate to control interdot tunnel coupling between the QDs~\cite{bonen2024investigation}. The two parallel-connected devices in the IC die differ from each other in terms of the dimensions of back-gate. In the selective back-gate (SB) device, the n$^{\mathrm{+}}$ doped silicon layer is limited to the dot gates region, whereas in the full back-gate (FB) device this doped silicon layer extends throughout the entire device geometry, as shown in Fig.~\ref{fig:fig_1}a. Switch gates (SW) are used to individually turn on or off the channel in the FB and SB devices, sharing a common drain terminal.

\section*{Monolithic IC Proposal}
Building upon the successful realization of quantized electron pumping with commercial foundry devices, we envision this technology to serve as a scalable quantum current standard based on the elementary charge~\cite{schoinas2024fast,norimoto2024statistical,dash2024parallelization}. The system architecture of the proposed IC, monolithically integrating one million charge pumps with on-chip classical control electronics on the FDSOI platform is shown in Fig.~\ref{fig:fig_1}b. This compact IC $(\approx~13.21~\mathrm{mm^2})$ operates at a temperature of 1.5~K, consuming approximately 12~mW of power—sufficiently low-power for the cooling capacity provided by commercially available cryostats at this temperature\cite{bluefors_xldhe, zpc_continuous_cold}. See Extended Data Table~\ref{tab:tab1} for a detailed breakdown of the power consumption and chip area requirements. 

The system can be divided into five main functional blocks: digital logic unit, dc bias control unit, rf drive control unit, switches and charge pumps (Fig.~\ref{fig:fig_1}b). The dc bias and rf drive control units generate variable dc voltages and phase-shifted rf waveforms based on digital bits provided by the digital logic unit. Considerable effort has been made to address electrical operation variability requirements across the large scale of industrially fabricated QDs during designing the on-chip dc and rf control circuitry. Only charge pumps compatible with our tunable parameters and operation protocol are selectively turned on, via digital inputs to CMOS inverter switches. The analogue dc and rf inputs to the parallel charge pumps are dynamically adjusted based on feedback from computing unit, to generate quantized current with the highest accuracy. See Methods Section: Proposed Monolithic IC Architecture for a detailed description. The summed quantized current from the parallelized charge pumps is collected at the common drain terminal and routed externally from the cryogenic IC to room-temperature for traceable measurement applications. When operated at 50~MHz, the upper bound of the total quantized current output is 8~$\mu\mathrm{A}$; the exact value depends on the uniformity of QD devices sharing common voltage tuning parameters. This enables a compact, table-top solution for realizing the SI ampere, traceable to the SI second.

\section*{Quantized electron pumping}
We pump single-electrons through the DQD in FB and SB devices by rapidly modulating $V_\mathrm{G1,\,FB/SB}$ and $V_\mathrm{G2,\,FB/SB}$ voltages with rf frequencies, $f_{\mathrm{FB/SB}}$. The following approach has been employed to tune the devices such that only one electron is transferred per frequency cycle across the DQD\cite{pothier1992single,connolly2013gigahertz,roche2013two}. First, we individually measure the common drain terminal current $(I_{D})$ as a function of dc voltages $V_\mathrm{G1,\,FB/SB}$ and $V_\mathrm{G2,\,FB/SB}$ with a common back-gate terminal voltage, $V_{\mathrm{BG}}$=3.5~V. The measured dc transport currents confirm the formation of DQDs in both the FB (Fig.~\ref{fig:fig_2}a) and SB (Fig.~\ref{fig:fig_2}b) devices. Owing to different back-gate dimensions, the FB and SB exhibits different electrical transport characteristics, when operated under same $V_{\mathrm{BG}}$ voltage. The FB device shows strong interdot coupling and lower channel resistance ($\approx92$~k$\Omega$), whereas the SB device displays weak interdot coupling and a higher channel resistance ($\approx4.2$~M$\Omega$). This attribute of higher channel resistance in the SB device primarily results from an increased threshold voltage in channel regions lacking back-gate coverage. As a result, the FB and SB devices posses different threshold voltage profiles. When operated at the same $V_{\mathrm{BG}} = 3.5$~V, they lie in different regimes of their respective transport characteristics, necessitating independent tuning of the top gates to achieve quantized electron pumping across the DQD system.

We locate the operation triple point within an appropriate range of $V_\mathrm{G1,\,FB/SB}$ and $V_\mathrm{G2,\,FB/SB}$ dot-gate voltages, where cotunneling events between the two QDs are suppressed and the bias triangles are well isolated with a background $I_\mathrm{D}\approx I_{\zeta}$, where $I_{\zeta}$ denotes the experimental noise floor. Figure~\ref{fig:fig_2}c shows a scan of $V_\mathrm{G1,\,SB}$ and $V_\mathrm{G2,\,SB}$ in this range (at small positive voltages), where the edges of the honeycomb pattern are less pronounced due to reduced cotunneling tunneling events. Dot gate dc voltages are fixed at the operation triple point (red plus, in Fig.~\ref{fig:fig_2}c), here at $V_\mathrm{{G1,\,SB}}$=109.0~mV and $V_\mathrm{{G2,\,SB}}$=120.5~mV. 

Finally, frequency-controlled quantized electron pumping is achieved by modulating the dot gate dc voltages with phase-shifted rf waveforms, $\widetilde{V}_{\mathrm{G1,\,SB}} \sin\left(2\pi f_{\mathrm{SB}} t + \phi_{\mathrm{G1,\,SB}}\right)$ and $\widetilde{V}_{\mathrm{G2,\,SB}} \sin\left(2\pi f_{\mathrm{SB}} t + \phi_{\mathrm{G2,\,SB}}\right)$, such that the resulting trajectory encircles the operation triple point~\cite{connolly2013gigahertz}. The elliptical trajectory (blue curve, in Fig.~\ref{fig:fig_2}c), also known as Lissajous plot, is determined by experimental parameters for $f_{\mathrm{SB}}$=10~MHz (see Extended Data Table \ref{tab:SB_pump_first}). The time-dependent electrostatic simulation of potential profile at four different time-stamps along the Lissajous trajectory during the first frequency cycle ($f_\mathrm{SB}$=10~MHz) of single electron pumping is shown in Fig.~\ref{fig:fig_2}d. Each pumping cycle involves: i. loading an electron into G2 QD from the drain reservoir $(N_{1},\,N_{2})_{t=0\, \mathrm{ns}}$; ii. capturing it in the G2 QD $(N_{1},\,N_{2}+1)_{t=20\, \mathrm{ns}}$; iii. shuttling the electron to the G1 QD $(N_{1}+1,\,N_{2})_{t=70\, \mathrm{ns}}$; and iv. unloading the electron to the source reservoir $(N_{1},\,N_{2})_{t=90\, \mathrm{ns}}$, yielding a positive pumped current when measured at the common drain terminal. 

The normalized pumped current $I_{D}/ef_\mathrm{FB/SB}$, ascertains the average number of pumped electrons, $\langle n \rangle$ per frequency cycle of the two phase-shifted rf waveforms\cite{dash2024silicon}. The observation of the $\langle n \rangle$=$1$ plateau as a function of $V_\mathrm{SD,\,SB}$ (Fig.~\ref{fig:fig_2}e) unambiguously validates single-electron pumping across the DQD in SB device\cite{jehl2013hybrid,connolly2013gigahertz}. Furthermore, Figure~\ref{fig:fig_2}g-h demonstrates bidirectional electron pumping as a function of $V_\mathrm{{G1,\,SB}}$ and $V_\mathrm{{G2,\,SB}}$ by applying 10~MHz rf waveforms with phase shifts of $\Delta \phi_{\mathrm{SB}}=90^\circ$ and $\Delta \phi_{\mathrm{SB}}=0^\circ$ phase shifted 10~MHz rf-waveforms around a triple point. The same tuning steps are repeated for the FB device (see Extended Data Table \ref{tab:FB_pump_first}) to realize single-electron pumping.

In order to maximize the quantized current generated from an individual charge pump, we gradually increase $f_\mathrm{FB}$ (Extended Data Fig.~\ref{fig:FB_pump}) and $f_\mathrm{SB}$ (Extended Data Fig.~\ref{fig:SB_pump}) up to a frequency of 50~MHz. Adiabatic pumping requires the modulation frequency to be small enough compared to the electron tunneling rate, ensuring sufficient time for the charge to get loaded, captured, shuttled, and unloaded across the DQD system. We calculated the upper bound of this $RC$ time constant to be approximately 49~ps for the FB device and 127~ps for the SB device. We observe that the frequency-dependent $I_\mathrm{D}$, measured at the flattest point of the $\langle n \rangle$=1 plateau for the FB (Fig.~\ref{fig:fig_3}a) and SB (Fig.~\ref{fig:fig_3}b) devices follows the linear $ef_\mathrm{FB/SB}$, expected for single-electron pumping.

\section*{Parallelization and accuracy measurement}
Next, we turn our attention towards demonstrating the scalability of the generated quantum current by simultaneously pumping single-electrons across both the FB and SB devices, resulting in a summed quantized current. In Fig.~\ref{fig:fig_4}a, we show the parallel operation of both the FB and SB charge pumping devices, with a parallel pump frequency, $f_{\parallel}$=50~MHz, yielding a total current of about 16~pA. We determine the accuracy of this summed quantized current, $I_{D}$=$I_\mathrm{P,\,FB}+I_\mathrm{P,\,SB}$ as a function of $f_{\parallel}$, using a high accuracy measurement scheme~\cite{giblin2012towards,zhao2017thermal,rossi2014accurate,bae2015precision,giblin2020realisation,giblin2023precision,yamahata2016gigahertz}. 

In the accuracy measurement (Fig.~\ref{fig:fig_4}a), we combine the $I_\mathrm{P,\,FB}+I_\mathrm{P,\,SB}$ with a room-temperature generated reference current of opposite polarity, $I_\mathrm{R}\equiv -V_\mathrm{R}/R_\mathrm{R}$~\cite{giblin2012towards,rossi2014accurate}. This $I_\mathrm{R}$ is traceable to primary voltage and resistance standards, with relative uncertainties $u_{S,\,V,}$ and $u_{S,\,\Omega}$, respectively (see Extended Data Table~\ref{tab:uncertainty-budget}). The combined resultant null current, $I_\mathrm{null}\equiv I_\mathrm{P,\,FB}+I_\mathrm{P,\,SB}-I_\mathrm{R}$ signifies the deviation of the pumped current from the SI traceable reference current. The contribution of experimental setup drift and offset currents on accuracy evaluation is alleviated through synchronous switching between $I_\mathrm{null}$ and $I_\mathrm{zero}$. Additionally, long averaging over 2500 data points reduces the system noise, contributing to the Type A uncertainty. A single raw data trace measured at $f_{\parallel}$=50~MHz over two alternating on/off cycles is shown in Fig.~\ref{fig:fig_4}b.

The accuracy of the parallel pump, is given as $|\Delta I/ef_{\parallel}|$\cite{giblin2012towards}, where $\Delta I=\overline{I}_\mathrm{zero}^{\theta} -\overline{I}_\mathrm{null}^{\theta}$ is the difference between the corrected mean of zero and null currents. $\overline{I}_\mathrm{zero}^{\theta}=\overline{I}_\mathrm{zero}+I_\mathrm{\theta}$, where $I_{\theta}$ is the 0~V offset current determined by the voltage source calibration. $\overline{I}_\mathrm{null}^{\theta}=\overline{I}_\mathrm{null}-I_\mathrm{\epsilon}$, where $I_\mathrm{\epsilon}$ is the error in generated reference current due to voltage source resolution. The evaluated accuracy of the summed quantized current from the FB and SB devices, is added with the expanded uncertainty, $U$ (see Extended Data Table \ref{tab:uncertainty-budget} for a detailed description) to determine the total measurement uncertainty of the generated summed quantized current. The total relative uncertainty $U+ |\Delta I/ef_{\parallel}|$ along with its estimated standard deviation $\sigma_{I/ef_{\parallel}}$, as a function of $f_{\parallel}$ is summarized in Fig.~\ref{fig:fig_4}c. 

We calculated the lowest relative uncertainty of $(1.2 \pm 0.1) \times 10^{-3}$~A/A for $f_{\parallel}$=50~MHz. This uncertainty figure of merit is orders of magnitude higher than that achieved by state-of-the-art single-electron pumps\cite{zhao2017thermal, giblin2012towards, giblin2020realisation}, in part due to limitations imposed by the measurement instrumentation and setup capabilities~\cite{giblin2020realisation, giblin2023precision}. It is worth noting that, although the widest quantized plateaus were observed in both the FB and SB devices for $f_\mathrm{FB/SB}$=10~MHz, the parallel pump architecture exhibited its lowest accuracy of $(9.8 \pm 0.5) \times 10^{-3}$~A/A at this frequency. This counterintuitive result suggests possible degree of randomness in the accuracy measurement setup and/or instability of the QD devices during the long-duration accuracy measurements where $f_{\parallel}$ was varied from 50~MHz to 10~MHz. 

Charge noise measurements (see Extended Data Fig.~\ref{fig:charge_noise}) indicate that both FB and SB devices exhibit higher charge noise at lower frequency scales (on the order of $10^{-1}$~Hz, corresponding to tens of seconds). Further, extrapolating the electrochemical potential noise spectra to longer timescales (on the order of minutes), based on the power-law fit exponent $\beta$ (see Extended Data Table~\ref{tab:noise-fit}), indicates that the FB device is more susceptible to low-frequency charge noise than the SB device. This characteristic is also evident from the shift in the charge stability map shown in Fig.~\ref{fig:fig_1}a,  as well as the repeated need to retune the dot-gate voltages in the FB device to maintain proximity to the operational triple point of the DQD system (see Extended Data Tables~\ref{tab:FB_pump_first} and~\ref{tab:FB_pump_second}). Before the null current measurements which were recorded over approximately 160 minutes the tuning parameters were individually optimized for the FB and SB devices. This involved first setting the dc operating triple point to the optimal bias condition, followed by determining the optimal  $V_\mathrm{SD}^{*}$ (see Extended Data Fig.~\ref{fig:FB_pump} and ~\ref{fig:SB_pump}) for each $f_{\parallel}$. However, these optimal voltages may have drifted over the long measurement period, potentially causing a deviation from the flattest point on the quantized current plateau. In future, implementing dynamic voltage and phase feedback would help maintain long-term charge-pump stability and ensure reliable quantization accuracy.

\section*{Discussion}
State-of-the-art quantum current sources generate SI traceable current standard with up to three orders of magnitude accuracy better than calibration and measurement capabilities, albeit over a narrow operating range\cite{djordjevic2025primary,kaneko2023perspectives, shaikhaidarov2022quantized, kaap2024demonstration}. While charge pumps offer sub-ppm accuracy up to hundreds of picoamperes, Ohm’s law-based quantum generators\cite{djordjevic2025primary, brun2016practical, chae2022series} yield currents only above microampere level. Our demonstration of quantum current generation in a commercial CMOS foundry fabricated QD device provides compelling evidence for accurate current realization across a wide dynamic range through practical large-scale integration.

This research marks one of the earliest applied efforts in translating low-temperature quantum effect device concept from scientific research into an industrial context. The clock-controlled, on-demand charge-transfer characteristics in the CMOS QD array is an important milestone not only for quantum metrology, but also for the semiconductor spin qubit community. As such CMOS-compatible architectures could be extended to enable spin information shuttling and routing across a scalable quantum processor\cite{kunne2024spinbus,de2024high}.

The charge noise in CMOS devices operated at cryogenic temperatures remains relatively high in their present state. This stems from the $\mathrm{HfO}_{2}$ high-$k$ metal gate stack, featuring a much thinner gate oxide in both FB and SB device structures. In contrast, more recent QD arrays fabricated using the 22FDX\textsuperscript{\textregistered} process, with thicker gate oxides, are expected to exhibit significantly lower charge noise. A lower charge noise is critical not only for generating precise quantum currents but also for enabling high-fidelity quantum applications such as spin qubits, aiming for translating academic and research foundry knowledge to industrial-grade CMOS QD devices\cite{steinacker2024300, connors2022charge}.

Our monolithic IC design, featuring one million parallel-connected charge-pumps, offers a promising path toward scalable quantum current generation, as suggested by chip area and power consumption estimates (Extended Data Table~\ref{tab:tab1}). While tuning the QD devices in this study required intensive manual effort and scientific knowledge, the process could be simplified using automated measurement frameworks based on machine learning algorithms and dynamic feedbacks~\cite{schoinas2024fast, thomas2025rapid, neyens2024probing}. Collectively, this lays the foundation for a compact, portable, and accurate quantum-based current standard—contributing toward a vision of realizing the SI standards on-chip~\cite{tzalenchuk2022expanding}.






\section{Methods}
\setcounter{section}{0}
\subsection*{Transport measurement}\label{methods:measurement_setup}
The cryogenic measurements were performed in an ICE Oxford pumped $^{4}\mathrm{He}$ system, with the IC enclosure mounted on the cold finger, maintaining a base temperature of 1.5~K. The sample space temperature was monitored using a Lakeshore 336 temperature controller. The dc voltages and rf waveforms are generated using a QDevil QDAC-\text{II} precision voltage source and a Tektronix AWG5200, respectively. 

The dc voltages were supplied to switch gates, source ohmic leads and common back-gate terminal via NEWT lines, filtered at 150~MHz using a cryogenic low-pass filter. The rf waveforms were attenuated by 20~dB at room temperature and combined with the dc voltages using SHF BT45R-B broadband bias tees. The combined dc-rf signal was routed to dot gates through SMA lines with a bandwidth 0 to 3~GHz. The drain current was amplified using a room-temperature transimpedance amplifier (FEMTO DLPCA-200) with a gain of $A_{V/I}=10^{9}$~V/A, followed by a voltage preamplifier (SIM910 JFET) with a gain of $A_{V}=10$. The voltage preamplifier was used primarily to avoid ground loop formation. The voltage signal was then filtered through a low-pass filter with a cutoff frequency of $f_{c}=45$~Hz, to suppress coupled high-frequency noise, and finally measured using a Keysight 34410A digital multimeter (DMM).

\subsection{Accuracy measurement}\label{methods:accuracy_setup}
The  QDevil QDAC-\text{II} high-precision low-noise dc voltage source was calibrated using a Keysight 3458A DMM. This DMM was itself calibrated against an electronic voltage reference (VS4, Statronics), which is traceable to the Josephson voltage standard maintained by the National Measurement Institute Australia. The 1~G$\Omega$ resistor (Guildline 65206) was calibrated using a Fluke 8508A DMM, with calibration traceable to the resistance standards maintained by the National Measurement Institute Australia. The generated null and/or zero current was measured using the transport measurement setup, where the Keysight 34410A DMM was replaced by the calibrated Keysight 3458A DMM. The uncertainty stated in this work has been calculated in accordance with the principles in JCGM 100:2008\cite{JCGM100_2008}, and gives an interval estimated to have a level of confidence of 95\%.

Commercial instruments are identified in this section only to adequately specify the experimental procedure. Such identification does not imply recommendation or endorsement by the National Measurement Institute, Australia, nor does it imply that the equipment identified is necessarily the best for the purpose.

\subsection*{Proposed Monolithic IC Architecture}\label{methods:IC_design}
The proposed monolithic IC architecture integrates one million parallel-connected CMOS-compatible QDs-based charge pumps along with their associated precise electrical control circuitry in 22-nm FDSOI platform. Dynamic digital inputs ($D_{\mathrm{SW}}$, $D_{\mathrm{DAC}}$, and $D_{\mathrm{VGA}}$) are supplied from a room-temperature computing unit to the on-chip digital logic unit. This unit includes three serial-in-parallel-out shift registers that control inputs to the CMOS switches, digital-to-analog converters\cite{vliex2020bias} (DACs) and variable-gain amplifiers\cite{huang2012ultra} (VGAs). The parallel-bit outputs $V_{\mathrm{CTRL,\,SW}}$, $V_{\mathrm{CTRL,\,DAC}}$, and $V_{\mathrm{CTRL,\,VGA}}$ of the shift registers are routed to the dc bias control unit, rf drive control unit, and CMOS switches, respectively.

The dc voltages critical to quantization accuracy of the charge pumps—such as dot gate voltages ($V_{\mathrm{G1}}$, $V_{\mathrm{G2}}$) and source-drain bias ($V_{\mathrm{SD}}$) are generated as the sum of an 8-bit coarse DAC $(X.1)$ and a 10-bit fine DAC $(X.2)$, providing 10~$\mu\mathrm{V}$ resolution over the working voltage range between 0 to 800~mV. Other control voltages, such as the back-gate ($V_{\mathrm{BG}}$) and switch gate ($V_{\mathrm{SW}}$), are set using 8-bit DACs with 15~mV resolution.

The rf drives are phase-synchronized to an input reference time-standard using a low-jitter phase-locked loop\cite{mansuri2003low} (PLL), which generates a stable sine wave. This signal is fed into a all-pass poly-phase shifter\cite{huang2012ultra} (PPS) to generate in-phase, $I = A \cdot \sin(\omega t)$ and quadrature, $Q = A \cdot \cos(\omega t)$ signals. These waveforms are independently conditioned by a pair of VGAs with gains $A_{V,\,I} = \gamma \cos(\phi)$ and $A_{V,\,Q} = \gamma \sin(\phi)$, controlled by digital-bits $V_{\mathrm{CTRL,\,VGA}} \langle 1:k\cdot16 \rangle$, to produce per-channel amplitude and phase shifted signal, $P = A \cdot {\gamma} \cdot \sin(\omega t + \phi)$. To enable independently tunable rf signals for one million charge pumps, a pair of 8-bit VGAs is dedicated to each cluster of 50 devices, resulting in $k+1 = 20{,}000$ independent rf channels. Each generated rf signal is ac coupled through a coupling capacitor to obtain $I_\mathrm{f}$ and $P_\mathrm{f} \langle1:k\rangle$ waveforms.

The filtered in-phase rf signal, $I_\mathrm{f}$ is superimposed on the dc gate voltage $V_{\mathrm{G1}}$ shared across one million charge pumps, to modulate the G1 QD potential. Given the likelihood of fabrication process variations across the large array of QDs, some devices may exhibit operational variability. To accommodate this, each phase-shifted rf signal $P_\mathrm{f}\langle1:k\rangle$ is individually added to $V_{\mathrm{G2}}$ for each 50-charge-pumps cluster, providing independently tuned amplitude and phase-shifted waveforms to modulate the G2 QD potential. 


The screening process begins by measuring the leakage transfer characteristics of each charge pump in a cluster, one at a time. During this step, all other charge pumps are turned off via the switch control line, $V_{\mathrm{CTRL,,SW}}\langle 1:m\rangle = 0$. These measurements reveal whether a device is functional and provide insight into threshold voltage variations. Based on the extracted transfer characteristics, appropriate dc voltage sweep ranges are defined for $V_{\mathrm{G1}}$ and $V_{\mathrm{G2}}$ for each DQD system\cite{dash2024parallelization}. Non-functional QD devices on the monolithic IC are selectively turned-off using SW gates located on both sides of the G1–G2 dot gates, independently controlled via $V_{\mathrm{CTRL,\, SW}}\langle 1:m\rangle$. Bias triangles are then recorded at locations exhibiting suppressed cotunneling and background current levels equivalent to measurement noise floor. This process is repeated for all 50 charge pumps in the cluster. Next, machine learning algorithms are employed to superimpose Lissajous curves onto the most suitable triple-point region of each bias triangle. A common rf amplitude and phase difference is determined for all the charge pumps in the cluster to maximize the number of operable charge pumps. This end-to-end process—from leakage tests to optimizing rf waveform parameters—is repeated across all 20,000 clusters to identify the most suitable common dc operating point, enabling quantized electron pumping in all functional charge pumps.

The summed quantized current ($I_\mathrm{D}$) from all operating charge pumps is collected at the common drain terminal and routed off-chip to room-temperature for realizing the quantum-based ``current standard''. However, as indicated by low-frequency charge noise spectra (Extended Data Fig. \ref{fig:charge_noise}) and shifts in voltage tuning parameters (Extended Data Tables \ref{tab:FB_pump_first} and \ref{tab:FB_pump_second}), maintaining long-term stability of the charge-pumps while providing reliable quantization accuracy would require dynamic voltage and phase feedback. A feedback switch monitors the output current in real-time and triggers the computing unit to update digital control bits—$D_{\mathrm{SW}}$, $D_{\mathrm{DAC}}$, and $D_{\mathrm{VGA}}$, which are then reloaded into the digital logic unit, completing the dynamic control loop. The parallel-out control voltages, including gain settings are updated through the shift register chains, allowing digital programmability while maintaining signal integrity.

\subsection*{Time-dependent electrostatic simulation}
Time-dependent electrostatic simulations in Fig.~\ref{fig:fig_2}d were carried out using the ac/dc Module of COMSOL Multiphysics. A three-dimensional finite element model of the FDSOI device was built within the software environment. The dc voltages and phase-shifted rf waveforms were then applied to the top-gate terminals, and the resulting electrostatic potentials were studied at four time-intervals within the first frequency cycle.

\subsection*{Charge noise spectroscopy}
Following the quantized electron pumping and accuracy measurements, the IC enclosure was transferred to a Bluefors LD400 dilution refrigerator system to measure charge noise of the FB and SB devices at 10~mK. This temperature was chosen to enable a fair comparison between the noise characteristics of commercial foundry (CF) fabricated FB and SB FDSOI QD devices studied in this work, and our previously measured QD devices fabricated using Academic cleanroom (AC) \cite{tanttu2024assessment} and research technology foundry (RTF) \cite{steinacker2024300} processes.

The noise spectra of the FB and SB devices were measured individually. A single QD was first defined under the G1 top-gate using similar voltage tuning parameters to those used in the charge-pumping experiments. The gate voltage $V_\mathrm{G1}$ was then fixed at a sensitive point of the Coulomb peak, and the conductance current was acquired from the common drain terminal as a function of lab time. The current was later converted to voltage using a Femto DLPCA-200 transimpedance amplifier, and then recorded with a PicoScope 4824A. Time-domain current noise data, $S_{I}(t)$ was then converted to voltage noise by using the relation, $S_{V}(t)=S_{I}(t)\cdot (dI_\mathrm{SD}/dV_\mathrm{G1})$, where $dI_\mathrm{SD}/dV_\mathrm{G1}$ is the slope of the Coulomb oscillation. The voltage noise spectrum is then Fourier transformed to obtain the single-sided frequency-domain noise power spectral density, $S_{V}(f)$. This was further converted into electrochemical potential noise using the relation $S_{\mu}(f) = \alpha \sqrt{S_{V}(f)}$, where $\alpha$ is the lever arm of the QD to G1 top-gate. The lever arms were determined to be $\alpha_\mathrm{G1-QD,\,FB}=0.35$~eV/V and $\alpha_\mathrm{G1-QD,\,SB}=0.72$~eV/V for the FB and SB devices, respectively. The charge noise spectra of the AC\cite{tanttu2024assessment} and RTF\cite{steinacker2024300} QD devices were acquired with a QM OPX$+$ and PicoScope 4824A, respectively.

\section*{Acknowledgements}
We acknowledge GlobalFoundries\textsuperscript{\texttrademark} for the chip fabrication through the 22FDX\textsuperscript{\textregistered} University Program and Nigel Cave for discussions. We thank National Measurement Institute, Australia for providing the voltage and resistance standards used for the accuracy measurements, Ilya Budovsky and Leigh Johnson for discussions. We thank Paul Steinacker and Nard Dumoulin Stuyck for sharing the charge noise data obtained from the RTF fabricated QD device\cite{steinacker2024300}. We thank Philip Mai and Jesus D. Cifuentes  for suggestions with the time-dependent electrostatic simulations. We thank Andrii Torgovkin for assistance with the cryogenic setup.

We acknowledge support from Australian Research Council (Grants No. DP200103515, and No. IM230100396), the U.S. Army Research Office (W911NF-17-1-0198), and the NSW Node of Australian National Fabrication Facility. A.D. and M.M.R. acknowledge scholarship support from the Sydney Quantum Academy, Australia. 

The views and conclusions contained in this document are those of the authors and should not be interpreted as representing the official policies, either expressed or implied, of the Army Research Office, the U.S. Air Force or the U.S. Government. The U.S. Government is authorized to reproduce and distribute reprints for Government purposes notwithstanding any copyright notation herein.

\section*{Author contributions}
A.D. performed the experiments and analysis with T.T.'s supervision, and inputs from D.G.. S.P.T. and S.B. designed the chip under S.P.V.'s supervision. A.D. and S.P.T. proposed the Monolithic IC architecture with S.P.V., A.S.D. and T.T.'s supervision. A.D. performed the time-dependent electrostatic simulation. D.G. calibrated the secondary voltage standard and assisted with accuracy measurements. D.G. and O.B. calibrated the resistance standard.  E.V. and T.T. performed the charge noise measurements with inputs from A.D.. S.Y. and K.W.C. packaged the chip. M.M.R., A.M., and A.L. contributed to the experimental hardware and cryogenic setup. A.D., S.P.T., D.G., M.K.F.,  E.V., J.Y.H., W.H.L., W.G., A.L., A.S., C.C.E., S.P.V., A.S.D. and T.T. participated in data interpretation. A.D. wrote the manuscript with contribution from all authors.

\section*{Corresponding authors}
Correspondence to A.D., or T.T..

\section*{Competing interests}
A.S.D. is the CEO and a director of Diraq Pty. Ltd.. M.K.F., S.Y., E.V., W.H.L., K.W.C., W.G., A.L., A.S., C.C.E., A.S.D. and T.T. declare equity interest in Diraq Pty. Ltd..

\section*{Data and Code availability}
The datasets and codes generated and/or analyzed during this study will be made available in an online repository.

\bibliographystyle{naturemag}
\bibliography{main.bib}

\begin{thebibliography}{10}
\expandafter\ifx\csname url\endcsname\relax
  \def\url#1{\texttt{#1}}\fi
\expandafter\ifx\csname urlprefix\endcsname\relax\def\urlprefix{URL }\fi
\providecommand{\bibinfo}[2]{#2}
\providecommand{\eprint}[2][]{\url{#2}}

\bibitem{tzalenchuk2022expanding}
\bibinfo{author}{Tzalenchuk, A.} \emph{et~al.}
\newblock \bibinfo{title}{The expanding role of national metrology institutes in the quantum era}.
\newblock \emph{\bibinfo{journal}{Nature Physics}} \textbf{\bibinfo{volume}{18}}, \bibinfo{pages}{724--727} (\bibinfo{year}{2022}).
\newblock \urlprefix\url{https://doi.org/10.1038/s41567-022-01659-z}.

\bibitem{kuramoto2022new}
\bibinfo{author}{Kuramoto, N.}
\newblock \bibinfo{title}{The new kilogram for new technology}.
\newblock \emph{\bibinfo{journal}{Nature Physics}} \textbf{\bibinfo{volume}{18}}, \bibinfo{pages}{720} (\bibinfo{year}{2022}).
\newblock \urlprefix\url{https://doi.org/10.1038/s41567-022-01615-x}.

\bibitem{stock2019revision}
\bibinfo{author}{Stock, M.}, \bibinfo{author}{Davis, R.}, \bibinfo{author}{de~Mirand{\'e}s, E.} \& \bibinfo{author}{Milton, M.~J.}
\newblock \bibinfo{title}{The revision of the si—the result of three decades of progress in metrology}.
\newblock \emph{\bibinfo{journal}{Metrologia}} \textbf{\bibinfo{volume}{56}}, \bibinfo{pages}{022001} (\bibinfo{year}{2019}).
\newblock \urlprefix\url{https://doi.org/10.1088/1681-7575/ab0013}.

\bibitem{BIPM_SI_2019}
\bibinfo{author}{{Bureau International des Poids et Mesures (BIPM)}}.
\newblock \emph{\bibinfo{title}{The International System of Units (SI)}} (\bibinfo{publisher}{BIPM}, \bibinfo{address}{Sèvres, France}, \bibinfo{year}{2019}), \bibinfo{edition}{9} edn.
\newblock \urlprefix\url{https://www.bipm.org/documents/20126/41483022/SI-Brochure-9-EN.pdf}.
\newblock \bibinfo{note}{Accessed: May, 2025}.

\bibitem{mohr2018data}
\bibinfo{author}{Mohr, P.~J.}, \bibinfo{author}{Newell, D.~B.}, \bibinfo{author}{Taylor, B.~N.} \& \bibinfo{author}{Tiesinga, E.}
\newblock \bibinfo{title}{Data and analysis for the codata 2017 special fundamental constants adjustment}.
\newblock \emph{\bibinfo{journal}{Metrologia}} \textbf{\bibinfo{volume}{55}}, \bibinfo{pages}{125--146} (\bibinfo{year}{2018}).
\newblock \urlprefix\url{https://doi.org/10.1088/1681-7575/aa99bc}.

\bibitem{schulte2020prospects}
\bibinfo{author}{Schulte, M.}, \bibinfo{author}{Lisdat, C.}, \bibinfo{author}{Schmidt, P.~O.}, \bibinfo{author}{Sterr, U.} \& \bibinfo{author}{Hammerer, K.}
\newblock \bibinfo{title}{Prospects and challenges for squeezing-enhanced optical atomic clocks}.
\newblock \emph{\bibinfo{journal}{Nature Communications}} \textbf{\bibinfo{volume}{11}}, \bibinfo{pages}{5955} (\bibinfo{year}{2020}).
\newblock \urlprefix\url{https://doi.org/10.1038/s41467-020-19403-7}.

\bibitem{dash2024silicon}
\bibinfo{author}{Dash, A.} \emph{et~al.}
\newblock \bibinfo{title}{Silicon-charge-pump operation limit above and below liquid-helium temperature}.
\newblock \emph{\bibinfo{journal}{Physical Review Applied}} \textbf{\bibinfo{volume}{21}}, \bibinfo{pages}{014040} (\bibinfo{year}{2024}).
\newblock \urlprefix\url{https://doi.org/10.1103/PhysRevApplied.21.014040}.

\bibitem{djordjevic2025primary}
\bibinfo{author}{Djordjevic, S.}, \bibinfo{author}{Behr, R.} \& \bibinfo{author}{Poirier, W.}
\newblock \bibinfo{title}{A primary quantum current standard based on the josephson and the quantum hall effects}.
\newblock \emph{\bibinfo{journal}{Nature Communications}} \textbf{\bibinfo{volume}{16}}, \bibinfo{pages}{1447} (\bibinfo{year}{2025}).
\newblock \urlprefix\url{https://doi.org/10.1038/s41467-025-56413-9}.

\bibitem{kaneko2023perspectives}
\bibinfo{author}{Kaneko, N.-H.}, \bibinfo{author}{Tanaka, T.} \& \bibinfo{author}{Okazaki, Y.}
\newblock \bibinfo{title}{Perspectives of the generation and measurement of small electric currents}.
\newblock \emph{\bibinfo{journal}{Measurement Science and Technology}} \textbf{\bibinfo{volume}{35}}, \bibinfo{pages}{011001} (\bibinfo{year}{2023}).
\newblock \urlprefix\url{https://doi.org/10.1088/1361-6501/ad03a2}.

\bibitem{fitzgerald2020next}
\bibinfo{author}{Fitzgerald, R.} \emph{et~al.}
\newblock \bibinfo{title}{The next generation of current measurement for ionization chambers}.
\newblock \emph{\bibinfo{journal}{Applied Radiation and Isotopes}} \textbf{\bibinfo{volume}{163}}, \bibinfo{pages}{109216} (\bibinfo{year}{2020}).
\newblock \urlprefix\url{https://doi.org/10.1016/j.apradiso.2020.109216}.

\bibitem{okazaki2022subfemtoampere}
\bibinfo{author}{Okazaki, Y.}, \bibinfo{author}{Tanaka, T.}, \bibinfo{author}{Saito, N.} \& \bibinfo{author}{Kaneko, N.-H.}
\newblock \bibinfo{title}{Subfemtoampere resolved ionization current measurements using a high-resistance transimpedance amplifier}.
\newblock \emph{\bibinfo{journal}{IEEE Transactions on Instrumentation and Measurement}} \textbf{\bibinfo{volume}{71}}, \bibinfo{pages}{1--8} (\bibinfo{year}{2022}).
\newblock \urlprefix\url{https://doi.org/10.1109/TIM.2022.3164155}.

\bibitem{feltin2009determination}
\bibinfo{author}{Feltin, N.} \& \bibinfo{author}{Piquemal, F.}
\newblock \bibinfo{title}{Determination of the elementary charge and the quantum metrological triangle experiment}.
\newblock \emph{\bibinfo{journal}{The European Physical Journal Special Topics}} \textbf{\bibinfo{volume}{172}}, \bibinfo{pages}{267--296} (\bibinfo{year}{2009}).
\newblock \urlprefix\url{https://doi.org/10.1140/epjst/e2009-01054-2}.

\bibitem{scherer2012quantum}
\bibinfo{author}{Scherer, H.} \& \bibinfo{author}{Camarota, B.}
\newblock \bibinfo{title}{Quantum metrology triangle experiments: a status review}.
\newblock \emph{\bibinfo{journal}{Measurement Science and Technology}} \textbf{\bibinfo{volume}{23}}, \bibinfo{pages}{124010} (\bibinfo{year}{2012}).
\newblock \urlprefix\url{https://doi.org/10.1088/0957-0233/23/12/124010}.

\bibitem{devoille2012quantum}
\bibinfo{author}{Devoille, L.} \emph{et~al.}
\newblock \bibinfo{title}{Quantum metrological triangle experiment at lne: measurements on a three-junction r-pump using a 20~000:1 winding ratio cryogenic current comparator}.
\newblock \emph{\bibinfo{journal}{Measurement Science and Technology}} \textbf{\bibinfo{volume}{23}}, \bibinfo{pages}{124011} (\bibinfo{year}{2012}).
\newblock \urlprefix\url{https://doi.org/10.1088/0957-0233/23/12/124011}.

\bibitem{scherer2019single}
\bibinfo{author}{Scherer, H.} \& \bibinfo{author}{Schumacher, H.~W.}
\newblock \bibinfo{title}{Single-electron pumps and quantum current metrology in the revised si}.
\newblock \emph{\bibinfo{journal}{Annalen der Physik}} \textbf{\bibinfo{volume}{531}}, \bibinfo{pages}{1800371} (\bibinfo{year}{2019}).
\newblock \urlprefix\url{https://doi.org/10.1002/andp.201800371}.

\bibitem{fujiwara2023silicon}
\bibinfo{author}{Fujiwara, A.}, \bibinfo{author}{Yamahata, G.}, \bibinfo{author}{Johnson, N.}, \bibinfo{author}{Nakamura, S.} \& \bibinfo{author}{Kaneko, N.}
\newblock \bibinfo{title}{Silicon quantum dot single-electron pumps for the closure of the quantum metrology triangle}.
\newblock \emph{\bibinfo{journal}{ECS Transactions}} \textbf{\bibinfo{volume}{112}}, \bibinfo{pages}{119} (\bibinfo{year}{2023}).
\newblock \urlprefix\url{https://doi.org/10.1149/11201.0119ecst}.

\bibitem{BIPM_Ampere_2019}
\bibinfo{title}{Mise en pratique for the definition of the ampere and other electrical units}.
\newblock \bibinfo{type}{Tech. Rep.} \bibinfo{number}{SI Brochure – 9th edition (2019) – Appendix 2}, \bibinfo{institution}{Bureau International des Poids et Mesures (BIPM)} (\bibinfo{year}{2019}).
\newblock \urlprefix\url{https://www.bipm.org/documents/20126/41489676/SI-App2-ampere.pdf}.
\newblock \bibinfo{note}{Accessed: May, 2025}.

\bibitem{giblin2020realisation}
\bibinfo{author}{Giblin, S.~P.} \emph{et~al.}
\newblock \bibinfo{title}{Realisation of a quantum current standard at liquid helium temperature with sub-ppm reproducibility}.
\newblock \emph{\bibinfo{journal}{Metrologia}} \textbf{\bibinfo{volume}{57}}, \bibinfo{pages}{025013} (\bibinfo{year}{2020}).
\newblock \urlprefix\url{https://doi.org/10.1088/1681-7575/ab72e0}.

\bibitem{giblin2023precision}
\bibinfo{author}{Giblin, S.~P.}, \bibinfo{author}{Yamahata, G.}, \bibinfo{author}{Fujiwara, A.} \& \bibinfo{author}{Kataoka, M.}
\newblock \bibinfo{title}{Precision measurement of an electron pump at 2 ghz; the frontier of small dc current metrology}.
\newblock \emph{\bibinfo{journal}{Metrologia}} \textbf{\bibinfo{volume}{60}}, \bibinfo{pages}{055001} (\bibinfo{year}{2023}).
\newblock \urlprefix\url{https://doi.org/10.1088/1681-7575/ace054}.

\bibitem{rossi2018gigahertz}
\bibinfo{author}{Rossi, A.} \emph{et~al.}
\newblock \bibinfo{title}{Gigahertz single-electron pumping mediated by parasitic states}.
\newblock \emph{\bibinfo{journal}{Nano Letters}} \textbf{\bibinfo{volume}{18}}, \bibinfo{pages}{4141--4147} (\bibinfo{year}{2018}).
\newblock \urlprefix\url{https://doi.org/10.1021/acs.nanolett.8b00874}.

\bibitem{kataoka2011tunable}
\bibinfo{author}{Kataoka, M.} \emph{et~al.}
\newblock \bibinfo{title}{Tunable nonadiabatic excitation in a single-electron quantum dot}.
\newblock \emph{\bibinfo{journal}{Physical Review Letters}} \textbf{\bibinfo{volume}{106}}, \bibinfo{pages}{126801} (\bibinfo{year}{2011}).
\newblock \urlprefix\url{https://doi.org/10.1103/PhysRevLett.106.126801}.

\bibitem{ahn2017upper}
\bibinfo{author}{Ahn, Y.-H.} \emph{et~al.}
\newblock \bibinfo{title}{Upper frequency limit depending on potential shape in a qd-based single electron pump}.
\newblock \emph{\bibinfo{journal}{Journal of Applied Physics}} \textbf{\bibinfo{volume}{122}} (\bibinfo{year}{2017}).
\newblock \urlprefix\url{https://doi.org/10.1063/1.5000319}.

\bibitem{johnson2024high}
\bibinfo{author}{Johnson, N.}, \bibinfo{author}{Yamahata, G.} \& \bibinfo{author}{Fujiwara, A.}
\newblock \bibinfo{title}{High frequency breakdown in quantum dots}.
\newblock \emph{\bibinfo{journal}{arXiv preprint arXiv:2410.08932}}  (\bibinfo{year}{2024}).
\newblock \urlprefix\url{https://doi.org/10.48550/arXiv.2410.08932}.

\bibitem{akmentinsh2025modeling}
\bibinfo{author}{Akmentinsh, A.}, \bibinfo{author}{Ubbelohde, N.} \& \bibinfo{author}{Kashcheyevs, V.}
\newblock \bibinfo{title}{Modeling shallow confinement in tunable quantum dots}.
\newblock \emph{\bibinfo{journal}{Physical Review B}} \textbf{\bibinfo{volume}{111}}, \bibinfo{pages}{075303} (\bibinfo{year}{2025}).
\newblock \urlprefix\url{https://doi.org/10.1103/PhysRevB.111.075303}.

\bibitem{stein2016robustness}
\bibinfo{author}{Stein, F.} \emph{et~al.}
\newblock \bibinfo{title}{Robustness of single-electron pumps at sub-ppm current accuracy level}.
\newblock \emph{\bibinfo{journal}{Metrologia}} \textbf{\bibinfo{volume}{54}}, \bibinfo{pages}{S1} (\bibinfo{year}{2016}).
\newblock \urlprefix\url{https://doi.org/10.1088/1681-7575/54/1/S1}.

\bibitem{yamahata2016gigahertz}
\bibinfo{author}{Yamahata, G.}, \bibinfo{author}{Giblin, S.~P.}, \bibinfo{author}{Kataoka, M.}, \bibinfo{author}{Karasawa, T.} \& \bibinfo{author}{Fujiwara, A.}
\newblock \bibinfo{title}{Gigahertz single-electron pumping in silicon with an accuracy better than 9.2 parts in 107}.
\newblock \emph{\bibinfo{journal}{Applied Physics Letters}} \textbf{\bibinfo{volume}{109}} (\bibinfo{year}{2016}).
\newblock \urlprefix\url{https://doi.org/10.1063/1.4953872}.

\bibitem{zhao2017thermal}
\bibinfo{author}{Zhao, R.} \emph{et~al.}
\newblock \bibinfo{title}{Thermal-error regime in high-accuracy gigahertz single-electron pumping}.
\newblock \emph{\bibinfo{journal}{Physical Review Applied}} \textbf{\bibinfo{volume}{8}}, \bibinfo{pages}{044021} (\bibinfo{year}{2017}).
\newblock \urlprefix\url{https://doi.org/10.1103/PhysRevApplied.8.044021}.

\bibitem{kim2022tuning}
\bibinfo{author}{Kim, B.-K.} \emph{et~al.}
\newblock \bibinfo{title}{Tuning current plateau regions in parallelized single-electron pumps}.
\newblock \emph{\bibinfo{journal}{AIP Advances}} \textbf{\bibinfo{volume}{12}}, \bibinfo{pages}{105118} (\bibinfo{year}{2022}).
\newblock \urlprefix\url{https://doi.org/10.1063/5.0117055}.

\bibitem{nakamura2023universality}
\bibinfo{author}{Nakamura, S.} \emph{et~al.}
\newblock \bibinfo{title}{Universality and multiplication of gigahertz-operated silicon pumps with parts per million-level uncertainty}.
\newblock \emph{\bibinfo{journal}{Nano Letters}} \textbf{\bibinfo{volume}{24}}, \bibinfo{pages}{9--15} (\bibinfo{year}{2023}).
\newblock \urlprefix\url{https://doi.org/10.1021/acs.nanolett.3c02858}.

\bibitem{schoinas2024fast}
\bibinfo{author}{Schoinas, N.} \emph{et~al.}
\newblock \bibinfo{title}{Fast characterization of multiplexed single-electron pumps with machine learning}.
\newblock \emph{\bibinfo{journal}{Applied Physics Letters}} \textbf{\bibinfo{volume}{125}} (\bibinfo{year}{2024}).
\newblock \urlprefix\url{https://doi.org/10.1063/5.0221387}.

\bibitem{norimoto2024statistical}
\bibinfo{author}{Norimoto, S.} \emph{et~al.}
\newblock \bibinfo{title}{Statistical study and parallelization of multiplexed single-electron sources}.
\newblock \emph{\bibinfo{journal}{Applied Physics Letters}} \textbf{\bibinfo{volume}{125}} (\bibinfo{year}{2024}).
\newblock \urlprefix\url{https://doi.org/10.1063/5.0225998}.

\bibitem{yamahata2025scalable}
\bibinfo{author}{Yamahata, G.}, \bibinfo{author}{Shimizu, T.}, \bibinfo{author}{Nishiguchi, K.} \& \bibinfo{author}{Fujiwara, A.}
\newblock \bibinfo{title}{Scalable parallel single-electron pumps in silicon with split-source control in the nanoampere regime}.
\newblock \emph{\bibinfo{journal}{arXiv preprint arXiv:2504.17273}}  (\bibinfo{year}{2025}).
\newblock \urlprefix\url{https://arxiv.org/abs/2504.17273}.

\bibitem{bonen2024investigation}
\bibinfo{author}{Bonen, S.}, \bibinfo{author}{Tripathi, S.~P.}, \bibinfo{author}{McIntosh, J.}, \bibinfo{author}{Jager, T.} \& \bibinfo{author}{Voinigescu, S.~P.}
\newblock \bibinfo{title}{Investigation of p-and n-type quantum dot arrays manufactured in 22nm fdsoi cmos at 2-4 k and 300 k}.
\newblock \emph{\bibinfo{journal}{IEEE Electron Device Letters}}  (\bibinfo{year}{2024}).
\newblock \urlprefix\url{https://doi.org/10.1109/LED.2024.3435380}.

\bibitem{zhao202380}
\bibinfo{author}{Zhao, J.} \& \bibinfo{author}{Voinigescu, S.~P.}
\newblock \bibinfo{title}{An 80-gbaud pam-4 gm-boosted variable-gain tia in 22-nm fdsoi}.
\newblock In \emph{\bibinfo{booktitle}{2023 IEEE BiCMOS and Compound Semiconductor Integrated Circuits and Technology Symposium (BCICTS)}}, \bibinfo{pages}{199--202} (\bibinfo{organization}{IEEE}, \bibinfo{year}{2023}).
\newblock \urlprefix\url{https://doi.org/10.1109/BCICTS54660.2023.10310939}.

\bibitem{bonen2018cryogenic}
\bibinfo{author}{Bonen, S.} \emph{et~al.}
\newblock \bibinfo{title}{Cryogenic characterization of 22-nm fdsoi cmos technology for quantum computing ics}.
\newblock \emph{\bibinfo{journal}{IEEE Electron Device Letters}} \textbf{\bibinfo{volume}{40}}, \bibinfo{pages}{127--130} (\bibinfo{year}{2018}).
\newblock \urlprefix\url{https://doi.org/10.1109/LED.2018.2880303}.

\bibitem{tripathi2022characterization}
\bibinfo{author}{Tripathi, S.~P.} \emph{et~al.}
\newblock \bibinfo{title}{Characterization and modeling of quantum dot behavior in fdsoi devices}.
\newblock \emph{\bibinfo{journal}{IEEE Journal of the Electron Devices Society}} \textbf{\bibinfo{volume}{10}}, \bibinfo{pages}{600--610} (\bibinfo{year}{2022}).
\newblock \urlprefix\url{https://doi.org/10.1109/JEDS.2022.3176205}.

\bibitem{amitonov2024commercial}
\bibinfo{author}{Amitonov, S.} \emph{et~al.}
\newblock \bibinfo{title}{Commercial cmos process for quantum computing: Quantum dots and charge sensing in a 22 nm fully depleted silicon-on-insulator process}.
\newblock \emph{\bibinfo{journal}{arXiv preprint arXiv:2412.08422}}  (\bibinfo{year}{2024}).
\newblock \urlprefix\url{https://doi.org/10.48550/arXiv.2412.08422}.

\bibitem{thomas2025rapid}
\bibinfo{author}{Thomas, E.~J.} \emph{et~al.}
\newblock \bibinfo{title}{Rapid cryogenic characterization of 1,024 integrated silicon quantum dot devices}.
\newblock \emph{\bibinfo{journal}{Nature Electronics}} \bibinfo{pages}{1--9} (\bibinfo{year}{2025}).
\newblock \urlprefix\url{https://doi.org/10.1038/s41928-024-01304-y}.

\bibitem{elbaz2025transport}
\bibinfo{author}{Elbaz, G.~A.} \emph{et~al.}
\newblock \bibinfo{title}{Transport characterization and quantum dot coupling in commercial 22fdx}.
\newblock \emph{\bibinfo{journal}{arXiv preprint arXiv:2501.10146}}  (\bibinfo{year}{2025}).
\newblock \urlprefix\url{https://doi.org/10.48550/arXiv.2501.10146}.

\bibitem{dash2024parallelization}
\bibinfo{author}{Dash, A.} \emph{et~al.}
\newblock \bibinfo{title}{Parallelization of charge-pumps in silicon for practical realization of the si ampere}.
\newblock In \emph{\bibinfo{booktitle}{2024 Conference on Precision Electromagnetic Measurements (CPEM)}}, \bibinfo{pages}{1--2} (\bibinfo{organization}{IEEE}, \bibinfo{year}{2024}).
\newblock \urlprefix\url{https://doi.org/10.1109/CPEM61406.2024.10646065}.

\bibitem{bluefors_xldhe}
\bibinfo{author}{{Bluefors}}.
\newblock \bibinfo{title}{{XLDHe High Power System}}.
\newblock \bibinfo{howpublished}{\url{https://bluefors.com}}.
\newblock \bibinfo{note}{Accessed: May, 2025}.

\bibitem{zpc_continuous_cold}
\bibinfo{author}{{Zero Point Cryogenics}}.
\newblock \bibinfo{title}{{Continuous Cold: The 1 K Upgradeable and Low Vibration Cryostat}}.
\newblock \bibinfo{howpublished}{\url{https://www.zpcryo.com}}.
\newblock \bibinfo{note}{Accessed: May, 2025}.

\bibitem{pothier1992single}
\bibinfo{author}{Pothier, H.}, \bibinfo{author}{Lafarge, P.}, \bibinfo{author}{Urbina, C.}, \bibinfo{author}{Esteve, D.} \& \bibinfo{author}{Devoret, M.~H.}
\newblock \bibinfo{title}{Single-electron pump based on charging effects}.
\newblock \emph{\bibinfo{journal}{Europhysics Letters}} \textbf{\bibinfo{volume}{17}}, \bibinfo{pages}{249} (\bibinfo{year}{1992}).
\newblock \urlprefix\url{https://doi.org/10.1209/0295-5075/17/3/011}.

\bibitem{connolly2013gigahertz}
\bibinfo{author}{Connolly, M.} \emph{et~al.}
\newblock \bibinfo{title}{Gigahertz quantized charge pumping in graphene quantum dots}.
\newblock \emph{\bibinfo{journal}{Nature Nanotechnology}} \textbf{\bibinfo{volume}{8}}, \bibinfo{pages}{417--420} (\bibinfo{year}{2013}).
\newblock \urlprefix\url{https://doi.org/10.1038/nnano.2013.73}.

\bibitem{roche2013two}
\bibinfo{author}{Roche, B.} \emph{et~al.}
\newblock \bibinfo{title}{A two-atom electron pump}.
\newblock \emph{\bibinfo{journal}{Nature Communications}} \textbf{\bibinfo{volume}{4}}, \bibinfo{pages}{1581} (\bibinfo{year}{2013}).
\newblock \urlprefix\url{https://doi.org/10.1038/ncomms2544}.

\bibitem{jehl2013hybrid}
\bibinfo{author}{Jehl, X.} \emph{et~al.}
\newblock \bibinfo{title}{Hybrid metal-semiconductor electron pump for quantum metrology}.
\newblock \emph{\bibinfo{journal}{Physical Review X}} \textbf{\bibinfo{volume}{3}}, \bibinfo{pages}{021012} (\bibinfo{year}{2013}).
\newblock \urlprefix\url{https://doi.org/10.1103/PhysRevX.3.021012}.

\bibitem{giblin2012towards}
\bibinfo{author}{Giblin, S.} \emph{et~al.}
\newblock \bibinfo{title}{Towards a quantum representation of the ampere using single electron pumps}.
\newblock \emph{\bibinfo{journal}{Nature communications}} \textbf{\bibinfo{volume}{3}}, \bibinfo{pages}{930} (\bibinfo{year}{2012}).
\newblock \urlprefix\url{https://doi.org/10.1038/ncomms1935}.

\bibitem{rossi2014accurate}
\bibinfo{author}{Rossi, A.} \emph{et~al.}
\newblock \bibinfo{title}{An accurate single-electron pump based on a highly tunable silicon quantum dot}.
\newblock \emph{\bibinfo{journal}{Nano letters}} \textbf{\bibinfo{volume}{14}}, \bibinfo{pages}{3405--3411} (\bibinfo{year}{2014}).
\newblock \urlprefix\url{https://doi.org/10.1021/nl500927q}.

\bibitem{bae2015precision}
\bibinfo{author}{Bae, M.-H.} \emph{et~al.}
\newblock \bibinfo{title}{Precision measurement of a potential-profile tunable single-electron pump}.
\newblock \emph{\bibinfo{journal}{Metrologia}} \textbf{\bibinfo{volume}{52}}, \bibinfo{pages}{195} (\bibinfo{year}{2015}).
\newblock \urlprefix\url{https://doi.org/10.1088/0026-1394/52/2/195}.

\bibitem{shaikhaidarov2022quantized}
\bibinfo{author}{Shaikhaidarov, R.~S.} \emph{et~al.}
\newblock \bibinfo{title}{Quantized current steps due to the a.c. coherent quantum phase-slip effect}.
\newblock \emph{\bibinfo{journal}{Nature}} \textbf{\bibinfo{volume}{608}}, \bibinfo{pages}{45--49} (\bibinfo{year}{2022}).
\newblock \urlprefix\url{https://doi.org/10.1038/s41586-022-04947-z}.

\bibitem{kaap2024demonstration}
\bibinfo{author}{Kaap, F.}, \bibinfo{author}{Kissling, C.}, \bibinfo{author}{Gaydamachenko, V.}, \bibinfo{author}{Gr{\"u}nhaupt, L.} \& \bibinfo{author}{Lotkhov, S.}
\newblock \bibinfo{title}{Demonstration of dual shapiro steps in small josephson junctions}.
\newblock \emph{\bibinfo{journal}{Nature Communications}} \textbf{\bibinfo{volume}{15}}, \bibinfo{pages}{8726} (\bibinfo{year}{2024}).
\newblock \urlprefix\url{https://doi.org/10.1038/s41467-024-53011-z}.

\bibitem{brun2016practical}
\bibinfo{author}{Brun-Picard, J.}, \bibinfo{author}{Djordjevic, S.}, \bibinfo{author}{Leprat, D.}, \bibinfo{author}{Schopfer, F.} \& \bibinfo{author}{Poirier, W.}
\newblock \bibinfo{title}{Practical quantum realization of the ampere from the elementary charge}.
\newblock \emph{\bibinfo{journal}{Physical Review X}} \textbf{\bibinfo{volume}{6}}, \bibinfo{pages}{041051} (\bibinfo{year}{2016}).
\newblock \urlprefix\url{https://doi.org/10.1103/PhysRevX.6.041051}.

\bibitem{chae2022series}
\bibinfo{author}{Chae, D.-H.}, \bibinfo{author}{Kim, M.-S.}, \bibinfo{author}{Oe, T.} \& \bibinfo{author}{Kaneko, N.-H.}
\newblock \bibinfo{title}{Series connection of quantum hall resistance array and programmable josephson voltage standard for current generation at one microampere}.
\newblock \emph{\bibinfo{journal}{Metrologia}} \textbf{\bibinfo{volume}{59}}, \bibinfo{pages}{065011} (\bibinfo{year}{2022}).
\newblock \urlprefix\url{https://doi.org/10.1088/1681-7575/ac97a0}.

\bibitem{kunne2024spinbus}
\bibinfo{author}{K{\"u}nne, M.} \emph{et~al.}
\newblock \bibinfo{title}{The spinbus architecture for scaling spin qubits with electron shuttling}.
\newblock \emph{\bibinfo{journal}{Nature Communications}} \textbf{\bibinfo{volume}{15}}, \bibinfo{pages}{4977} (\bibinfo{year}{2024}).
\newblock \urlprefix\url{https://doi.org/10.1038/s41467-024-49182-4}.

\bibitem{de2024high}
\bibinfo{author}{De~Smet, M.} \emph{et~al.}
\newblock \bibinfo{title}{High-fidelity single-spin shuttling in silicon}.
\newblock \emph{\bibinfo{journal}{Nature Nanotechnology}}  (\bibinfo{year}{2025}).
\newblock \urlprefix\url{https://doi.org/10.1038/s41565-025-01920-5}.

\bibitem{steinacker2024300}
\bibinfo{author}{Steinacker, P.} \emph{et~al.}
\newblock \bibinfo{title}{A 300 mm foundry silicon spin qubit unit cell exceeding 99\% fidelity in all operations}.
\newblock \emph{\bibinfo{journal}{arXiv preprint arXiv:2410.15590}}  (\bibinfo{year}{2024}).
\newblock \urlprefix\url{https://doi.org/10.48550/arXiv.2410.15590}.

\bibitem{connors2022charge}
\bibinfo{author}{Connors, E.~J.}, \bibinfo{author}{Nelson, J.}, \bibinfo{author}{Edge, L.~F.} \& \bibinfo{author}{Nichol, J.~M.}
\newblock \bibinfo{title}{Charge-noise spectroscopy of si/sige quantum dots via dynamically-decoupled exchange oscillations}.
\newblock \emph{\bibinfo{journal}{Nature Communications}} \textbf{\bibinfo{volume}{13}}, \bibinfo{pages}{940} (\bibinfo{year}{2022}).
\newblock \urlprefix\url{https://doi.org/10.1038/s41467-022-28519-x}.

\bibitem{neyens2024probing}
\bibinfo{author}{Neyens, S.} \emph{et~al.}
\newblock \bibinfo{title}{Probing single electrons across 300-mm spin qubit wafers}.
\newblock \emph{\bibinfo{journal}{Nature}} \textbf{\bibinfo{volume}{629}}, \bibinfo{pages}{80--85} (\bibinfo{year}{2024}).
\newblock \urlprefix\url{https://doi.org/10.1038/s41586-024-07275-6}.

\bibitem{JCGM100_2008}
\bibinfo{title}{Evaluation of measurement data -- guide to the expression of uncertainty in measurement}.
\newblock \bibinfo{type}{Tech. Rep.} \bibinfo{number}{JCGM 100:2008}, \bibinfo{institution}{Joint Committee for Guides in Metrology (JCGM)} (\bibinfo{year}{2008}).
\newblock \urlprefix\url{https://www.bipm.org/documents/20126/2071204/JCGM_100_2008_E.pdf}.
\newblock \bibinfo{note}{Accessed: May, 2025}.

\bibitem{vliex2020bias}
\bibinfo{author}{Vliex, P.} \emph{et~al.}
\newblock \bibinfo{title}{Bias voltage dac operating at cryogenic temperatures for solid-state qubit applications}.
\newblock \emph{\bibinfo{journal}{IEEE solid-state circuits letters}} \textbf{\bibinfo{volume}{3}}, \bibinfo{pages}{218--221} (\bibinfo{year}{2020}).
\newblock \urlprefix\url{https://doi.org/10.1109/LSSC.2020.3011576}.

\bibitem{huang2012ultra}
\bibinfo{author}{Huang, Y.-Y.} \emph{et~al.}
\newblock \bibinfo{title}{An ultra-compact, linearly-controlled variable phase shifter designed with a novel rc poly-phase filter}.
\newblock \emph{\bibinfo{journal}{IEEE transactions on microwave theory and techniques}} \textbf{\bibinfo{volume}{60}}, \bibinfo{pages}{301--310} (\bibinfo{year}{2012}).
\newblock \urlprefix\url{https://doi.org/10.1109/TMTT.2011.2177856}.

\bibitem{mansuri2003low}
\bibinfo{author}{Mansuri, M.} \& \bibinfo{author}{Yang, C.-K.}
\newblock \bibinfo{title}{A low-power adaptive bandwidth pll and clock buffer with supply-noise compensation}.
\newblock \emph{\bibinfo{journal}{IEEE Journal of Solid-State Circuits}} \textbf{\bibinfo{volume}{38}}, \bibinfo{pages}{1804--1812} (\bibinfo{year}{2003}).
\newblock \urlprefix\url{https://doi.org/10.1109/JSSC.2003.818300}.

\bibitem{tanttu2024assessment}
\bibinfo{author}{Tanttu, T.} \emph{et~al.}
\newblock \bibinfo{title}{Assessment of the errors of high-fidelity two-qubit gates in silicon quantum dots}.
\newblock \emph{\bibinfo{journal}{Nature Physics}} \bibinfo{pages}{1--6} (\bibinfo{year}{2024}).
\newblock \urlprefix\url{https://doi.org/10.1038/s41567-024-02614-w}.

\end{thebibliography}

\clearpage
\newpage
\onecolumngrid
\vfill

\section*{Extended data}

\setcounter{figure}{0}
\setcounter{table}{0}

\captionsetup[figure]{name={\bf{Extended Data Fig.}},labelsep=line,justification=centerlast,font=small}

\captionsetup[table]{name={\bf{Extended Data Table}},labelsep=period,justification=justified,font=small}

\hspace{5cm}
\begin{table*}[hbt!]
    \centering
    \renewcommand{\arraystretch}{1}
    \setlength{\tabcolsep}{8 pt}
    \caption{Power consumption and chip size area breakdown for an integrated circuit containing one million charge pumps and on-chip control electronics.}
    \label{tab:tab1}
    \begin{tabular}{l c c c}
        \toprule
        \textbf{Component} & \textbf{Units required} & \textbf{Power (mW)} & \textbf{Area (\textbf{\textrm{mm}}$^{\mathbf{2}}$)} \\
        \midrule
        \addlinespace
        \multicolumn{1}{c}{\textsc{\footnotesize Charge pumps}} \\
        \noindent Quantum dot array & \(1.00 \times 10^{6}\) & \(1.60 \times 10^{-4}\) & \(3.70 \times 10^{-1}\) \\
        \addlinespace
        \addlinespace
        \addlinespace
        \multicolumn{1}{c}{\textsc{\footnotesize Switch}} \\
        \noindent CMOS inverter & \(1.00 \times 10^{6}\) & \(2.00 \times 10^{-3}\) & \(3.40 \times 10^{-1}\) \\
        \addlinespace
        \addlinespace
        \addlinespace
        \multicolumn{1}{c}{\textsc{\footnotesize RF Drive Control Unit}} \\
        \addlinespace
         \noindent Phase Locked Loop (PLL) & \(1.00 \times 10^{0}\) & \(1.00 \times 10^{1}\) & \(1.00 \times 10^{-1}\) \\
         \noindent Poly Phase Shifter (PPS) & \(1.00\times 10^{0}\) & \(1.00 \times 10^{0}\) & \(4.80 \times 10^{-2}\) \\
         \noindent Voltage Controlled Oscillator (VCO) & \(1.00\times 10^{0}\) & \(5.00 \times 10^{-1}\) & \(4.00 \times 10^{-2}\) \\      
         \noindent Variable Gain Amplifier (VGA) & \(4.00 \times 10^{4}\) & \(3.20 \times 10^{-5}\) & \(1.60\times 10^{0}\) \\   
         \noindent Capacitor & \(2.00 \times 10^{4}\) & {-} & \(4.00\times 10^{0}\) \\
        \addlinespace
        \addlinespace
        \addlinespace
        \multicolumn{1}{c}{\textsc{\footnotesize DC Bias Control Unit}} \\
        \addlinespace
         \noindent Digital to Analog Converter (DAC) & \(8.00\times 10^{0}\) & \(1.60 \times 10^{-2}\) & \(2.00 \times 10^{-2}\) \\
        \addlinespace
        \addlinespace
        \addlinespace
        \multicolumn{1}{c}{\textsc{\footnotesize Digital Logic Unit}} \\
        \addlinespace
         \noindent Shift register bits$-$Switch & \(1.00 \times 10^{6}\) & \(2.00 \times 10^{-2}\) & \(3.40 \times 10^{0}\) \\       
         \noindent Shift register bits$-$DAC & \(7.00 \times 10^{1} \) & \(1.40 \times 10^{-6}\) & \(2.44 \times 10^{-4}\) \\      
         \noindent Shift register bits$-$VGA & \(3.20 \times 10^{5}\) & \(6.40 \times 10^{-3}\) & \(1.09 \times 10^{0}\) \\

        \addlinespace
        \addlinespace
        \addlinespace
        \multicolumn{1}{c}{\textsc{\footnotesize Placement and Routing}} 
    & {-} & {-} & \(2.20 \times 10^{0}\) \\
        \addlinespace
        \midrule
        \textbf{TOTAL} & {} & \(\mathbf{11.54} \) & \(\mathbf{13.21} \) \\
        \bottomrule
    \end{tabular}
\end{table*}

\newpage

\begin{figure*}[ht!]
    \includegraphics[width=0.6\textwidth,angle = 0]{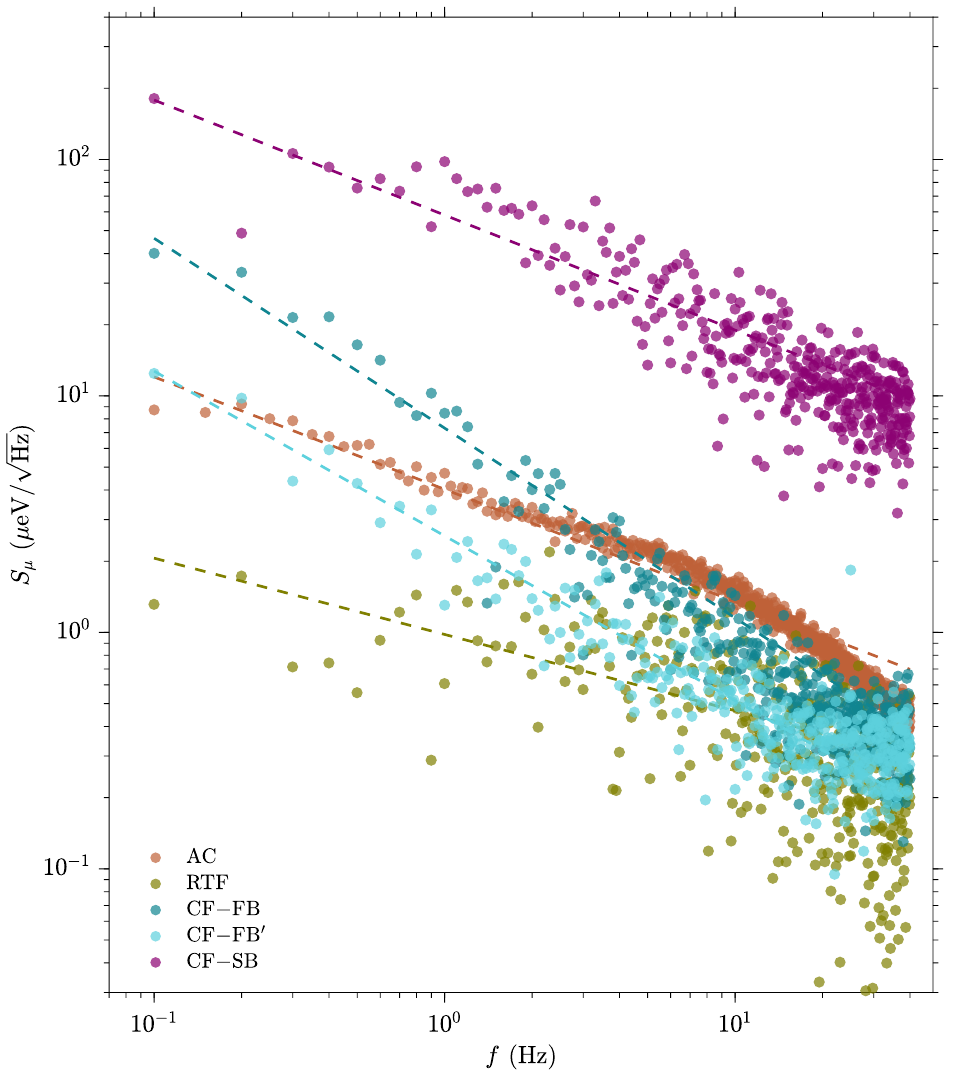}
    \caption{\textbf{Charge Noise Spectrum (T=10 mK).} Electrochemical potential noise spectra of quantum dot devices fabricated in the Academic cleanroom (AC), Research Technology Foundry (RTF), and Commercial Foundry (CF). The charge noise was recorded at a sensitive point of the Coulomb peak. The AC and RTF noise spectra were measured from Device A in Ref.~\cite{tanttu2024assessment} and Ref.~\cite{steinacker2024300}, respectively. The CF$-$FB and CF$-$SB devices are the full and selective back-gate devices used in this work. CF$-$FB and CF$-$FB$^\prime$ noise spectra were recorded from the same full back-gate device about one hour apart. The dashed lines represents noise spectrum fit to power-law model, $S_{\mu}(f) = A/f^{\frac{\beta}{2}}$.} 
    \label{fig:charge_noise}
\end{figure*}

\begin{table}[htbp]
  \centering
  \renewcommand{\arraystretch}{1.3}
  \setlength{\tabcolsep}{15pt}
  \caption{Charge noise at 1~Hz $(A)$ and power‐law fit exponent $(\beta)$ of the noise spectra.}
  \label{tab:noise-fit}
  \begin{tabular}{lccc}
    \toprule
    Quantum dot device 
      & \(A\) \(\bigl(\mu\mathrm{eV}/\sqrt{\mathrm{Hz}}\bigr)\) 
      & \(\beta\)
      & Reference \\
    \midrule
    \(\mathrm{Academic\,Cleanroom}\)                            
      &  4.03 & 0.95 & \cite{tanttu2024assessment} \\
    \(\mathrm{Research\,Technology\,Foundry}\)                      
      &  0.98 & 0.65 & \cite{steinacker2024300} \\
    \(\mathrm{Commercial\,Foundry}\text{-}\mathrm{FB}\)            
      &  7.31 & 1.61 &  This work\\
    \(\mathrm{Commercial\,Foundry}\text{-}\mathrm{FB}^{\prime}\)   
      &  2.56 & 1.39 &  This work\\
    \(\mathrm{Commercial\,Foundry}\text{-}\mathrm{SB}\)            
      & 58.18 & 0.97 &  This work\\
    \bottomrule
  \end{tabular}
\end{table}

\newpage

\begin{figure*}[ht!]
    \includegraphics[angle = 0]{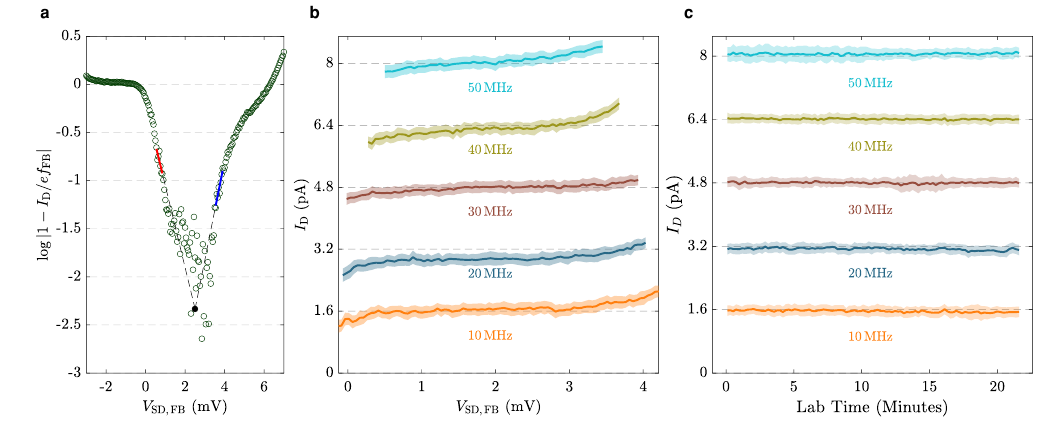}
    \caption{\textbf{
    Full back-gate device biasing and stability of pumped current (T=1.5K).} \textbf{a}, The $\log|1 - I_{\mathrm{D}} / ef_{\mathrm{FB}}|$ plot as a function of source bias $(V_{\mathrm{SD,\,FB}})$ at $f_{\mathrm{FB}}$=20~MHz. The red-blue lines represent linear fits to the data at the edges of the quantized current plateau. The black dashed lines show the extrapolated fits used to determine the optimal source bias for single-electron pumping ($V_{\mathrm{SD,\,FB}}^{*}$), labeled as black circle. Frequency-dependent average pumped current, $I_{\mathrm{D}}$ with error bounds as a function of \textbf{b}, $V_{\mathrm{SD,\,FB}}$, and \textbf{c}, lab time. See Extended Data Table \ref{tab:FB_pump_first} for values of $V_{\mathrm{SD,\,FB}}^{*}$ used to measure frequency-dependent $I_{\mathrm{D}}$ as a function of lab time. Error bounds are calculated by averaging 100 data points, each measured with an integration time of 20~ms and a wait time of 100~ms. In \textbf{b}, the data are horizontally shifted for clarity.
    }
    \label{fig:FB_pump}
\end{figure*}

\begin{table}[ht]
  \centering
  \renewcommand{\arraystretch}{1.12}
  \setlength{\tabcolsep}{7 pt}
  \caption{Full back-gate device ac and dc parameters used to realize quantized electron pumping.}
  \label{tab:FB_pump_first}
  \begin{tabular}{cccccccccc}
    \toprule
    \shortstack{$f_{\mathrm{FB}}$ \\ (MHz)} &
    \shortstack{$V_{\mathrm{BG}}$ \\ (V)} &
    \shortstack{$V_{\mathrm{SW,\,FB}}$ \\ (mV)} &
    \shortstack{$V_{\mathrm{SD,\,FB}}^{*}$ \\ (mV)} &
    \shortstack{$V_{\mathrm{G1,\,FB}}$ \\ (mV)} &
    \shortstack{$V_{\mathrm{G2,\,FB}}$ \\ (mV)} &
    \shortstack{$\widetilde{V}_{\mathrm{G1,\,FB}}$ \\ (mV)} &
    \shortstack{$\widetilde{V}_{\mathrm{G2,\,FB}}$ \\ (mV)} &
    \shortstack{$\phi_{\mathrm{G1,\,FB}}-\phi_{\mathrm{G2,\,FB}}$ \\ (degree)} \\
    \midrule
    10 & 3.5 & 250 & 1.75 & 50.6 & 65.5 & 12 & 12 & 140 \\
    20 & 3.5 & 250 & 2.25 & 50.6 & 65.5 & 13 & 13 & 140 \\
    30 & 3.5 & 250 & 2.35 & 50.6 & 65.5 & 14 & 14 & 120 \\
    40 & 3.5 & 250 & 2.85 & 50.6 & 65.5 & 15 & 15 & 110 \\
    50 & 3.5 & 250 & 2.90 & 50.6 & 65.5 & 16 & 16 & 110 \\
    \bottomrule
  \end{tabular}
\end{table}

\begin{table}[ht]
  \centering
  \renewcommand{\arraystretch}{1.12}
  \setlength{\tabcolsep}{10 pt}
  \caption{Full back-gate device ac and dc parameters used to realize quantized electron pumping, retuned two days after the initial measurements (Extended Data Table~\ref{tab:FB_pump_first}).}
  \label{tab:FB_pump_second}
  \begin{tabular}{ccccccccc}
    \toprule
    \shortstack{$f_{\mathrm{FB}}$ \\ (MHz)} &
    \shortstack{$V_{\mathrm{BG}}$ \\ (V)} &
    \shortstack{$V_{\mathrm{SW,\,FB}}$ \\ (mV)} &
    \shortstack{$V_{\mathrm{G1,\,FB}}$ \\ (mV)} &
    \shortstack{$V_{\mathrm{G2,\,FB}}$ \\ (mV)} &
    \shortstack{$\widetilde{V}_{\mathrm{G1,\,FB}}$ \\ (mV)} &
    \shortstack{$\widetilde{V}_{\mathrm{G2,\,FB}}$ \\ (mV)} &
    \shortstack{$\phi_{\mathrm{G1,\,FB}}-\phi_{\mathrm{G2,\,FB}}$ \\ (degree)} \\
    \midrule
    10 & 3.5 & 250 & 47.4 & 61.4 & 12 & 12 & 150 \\
    20 & 3.5 & 250 & 47.4 & 61.4 & 13 & 13 & 140 \\
    30 & 3.5 & 250 & 47.4 & 61.4 & 14 & 14 & 120 \\
    40 & 3.5 & 250 & 47.4 & 61.4 & 15 & 15 & 110 \\
    50 & 3.5 & 250 & 47.4 & 61.4 & 16 & 16 & 110 \\
    \bottomrule
  \end{tabular}
\end{table}
\newpage

\begin{figure*}[ht!]
    \includegraphics[angle = 0]{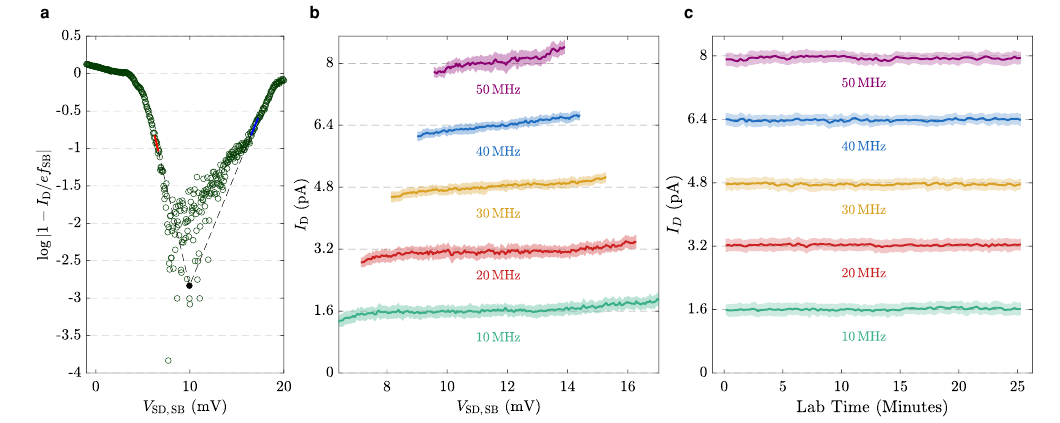}
    \caption{\textbf{
    Selective back-gate device biasing and stability of pumped current (T=1.5K).} \textbf{a}, The $\log|1 - I_{\mathrm{D}} / ef_{\mathrm{SB}}|$ plot as a function of source bias $(V_{\mathrm{SD,\,SB}})$ at $f_{\mathrm{SB}}$=20~MHz. The red-blue lines represent linear fits to the data at the edges of the quantized current plateau. The black dashed lines show the extrapolated fits used to determine the optimal source bias for single-electron pumping ($V_{\mathrm{SD,\,SB}}^{*}$), labeled as black circle. Frequency-dependent average pumped current, $I_{\mathrm{D}}$ with error bounds as a function of \textbf{b}, $V_{\mathrm{SD,\,SB}}$, and \textbf{c}, lab time. See Extended Data Table \ref{tab:SB_pump_first} for values of $V_{\mathrm{SD,\,SB}}^{*}$ used to measure frequency-dependent $I_{\mathrm{D}}$ as a function of lab time. Error bounds are calculated by averaging 100 data points, each measured with an integration time of 20~ms and a wait time of 100~ms. In \textbf{b}, the data are horizontally shifted for clarity.
    }
    \label{fig:SB_pump}
\end{figure*}

\begin{table}[ht]
  \centering
  \renewcommand{\arraystretch}{1.12}
  \setlength{\tabcolsep}{7 pt}
  \caption{Selective back-gate device ac and dc parameters used to realize quantized electron pumping.}
  \label{tab:SB_pump_first}
  \begin{tabular}{cccccccccc}
    \toprule
    \shortstack{$f_{\mathrm{SB}}$ \\ (MHz)} &
    \shortstack{$V_{\mathrm{BG}}$ \\ (V)} &
    \shortstack{$V_{\mathrm{SW,\,SB}}$ \\ (mV)} &
    \shortstack{$V_{\mathrm{SD,\,SB}}^{*}$ \\ (mV)} &
    \shortstack{$V_{\mathrm{G1,\,SB}}$ \\ (mV)} &
    \shortstack{$V_{\mathrm{G2,\,SB}}$ \\ (mV)} &
    \shortstack{$\widetilde{V}_{\mathrm{G1,\,SB}}$ \\ (mV)} &
    \shortstack{$\widetilde{V}_{\mathrm{G2,\,SB}}$ \\ (mV)} &
    \shortstack{$\phi_{\mathrm{G1,\,SB}}-\phi_{\mathrm{G2,\,SB}}$ \\ (degree)} \\
    \midrule
    10 & 3.5 & 800 & 13.00 & 109.0 & 120.5 & 10 & 10 & -150 \\
    20 & 3.5 & 800 & 9.75 & 109.0 & 120.5 & 11 & 11 & -150 \\
    30 & 3.5 & 800 & 9.90 & 109.0 & 120.5 & 12 & 12 & -150 \\
    40 & 3.5 & 800 & 12.50 & 109.0 & 120.5 & 14 & 14 & -150 \\
    50 & 3.5 & 800 & 12.35 & 109.0 & 120.5 & 15 & 15 & -160 \\
    \bottomrule
  \end{tabular}
\end{table}

\begin{table}[ht]
  \centering
  \renewcommand{\arraystretch}{1.12}
  \setlength{\tabcolsep}{10 pt}
  \caption{Selective back-gate device ac and dc parameters used to realize quantized electron pumping, retuned two days after the initial measurements (Extended Data Table~\ref{tab:SB_pump_first}).}
  \label{tab:SB_pump_second}
  \begin{tabular}{ccccccccc}
    \toprule
    \shortstack{$f_{\mathrm{SB}}$ \\ (MHz)} &
    \shortstack{$V_{\mathrm{BG}}$ \\ (V)} &
    \shortstack{$V_{\mathrm{SW,\,SB}}$ \\ (mV)} &
    \shortstack{$V_{\mathrm{G1,\,SB}}$ \\ (mV)} &
    \shortstack{$V_{\mathrm{G2,\,SB}}$ \\ (mV)} &
    \shortstack{$\widetilde{V}_{\mathrm{G1,\,SB}}$ \\ (mV)} &
    \shortstack{$\widetilde{V}_{\mathrm{G2,\,SB}}$ \\ (mV)} &
    \shortstack{$\phi_{\mathrm{G1,\,SB}}-\phi_{\mathrm{G2,\,SB}}$ \\ (degree)} \\
    \midrule
    10 & 3.5 & 800 & 109.0 & 120.5 & 10 & 10 & -150 \\
    20 & 3.5 & 800 & 109.0 & 120.5 & 11 & 11 & -150 \\
    30 & 3.5 & 800 & 109.0 & 120.5 & 12 & 12 & -150 \\
    40 & 3.5 & 800 & 109.0 & 120.5 & 14 & 14 & -150 \\
    50 & 3.5 & 800 & 109.0 & 120.5 & 15 & 15 & -160 \\
    \bottomrule
  \end{tabular}
\end{table}


\setcounter{figure}{0}

\clearpage

\begin{table}[htbp]
\centering
  \renewcommand{\arraystretch}{1.5}
  \setlength{\tabcolsep}{8pt} 

\caption{Uncertainty budget breakdown as a function of $f_{\parallel}$ for the accuracy measurements. All reported
uncertainties are expressed in units of A/A. The value of $u_{S,\,\Omega}$ is calculated by accounting for contributions from the calibration, resolution, and transfer uncertainty of the Fluke 8508A DMM, as well as the calibration, drift, and temperature coefficient of the 1~$\mathrm{G}\Omega$ reference resistor. Similarly, the value of $u_{S,\,V}$ includes the calibration, drift, and temperature coefficient of the Statronics VS4 electronic voltage reference, along with the resolution and stability of the Keysight 3458A DMM, and the resolution and drift of the QDevil QDAC-\text{II}. The combined uncertanity is given by: $u_{C}=\sqrt{u_{S,\,\Omega}^{2}+u_{S,\,V}^{2}+u_{A}^{2}+u_{E}^{2}}$. Finally, the expanded uncertainty $U$ is calculated by applying a coverage factor $k$ corresponding to a 95\% confidence level\cite{JCGM100_2008}.}

\label{tab:uncertainty-budget}
\begin{tabular}{l@{\hskip 6pt}cccccc}
\toprule
\multicolumn{2}{c}{} & \multicolumn{5}{c}{$f_{\parallel}$ (MHz)} \\
\cmidrule(lr){3-7}
 &  & 10 & 20 & 30 & 40 & 50 \\
\midrule
Reference resistor calibration & $u_{S,\,\Omega}$ 
  & $5.0\times10^{-5}$ & $5.0\times10^{-5}$ 
  & $5.0\times10^{-5}$ & $5.0\times10^{-5}$ 
  & $5.0\times10^{-5}$ \\
Reference voltage source calibration & $u_{S,\,V}$ 
  & $3.5\times10^{-4}$ & $2.0\times10^{-4}$ 
  & $1.5\times10^{-4}$ & $1.0\times10^{-4}$ 
  & $1.0\times10^{-4}$ \\
Transimpedance amplifier converter & $u_{A}$ 
  & $1.0\times10^{-4}$ & $1.0\times10^{-4}$ 
  & $1.0\times10^{-4}$ & $1.0\times10^{-4}$ 
  & $1.0\times10^{-4}$ \\
Type A ESDM measurement & $u_{E}$ 
  & $1.3\times10^{-3}$ & $7.0\times10^{-4}$ 
  & $5.0\times10^{-4}$ & $4.0\times10^{-4}$ 
  & $3.0\times10^{-4}$ \\

Combined standard uncertainty & $u_{C}$ 
  & $1.4\times10^{-3}$ & $8.0\times10^{-4}$ 
  & $6.0\times10^{-4}$ & $5.0\times10^{-4}$ 
  & $4.0\times10^{-4}$ \\
  
  
\midrule
\textbf{Expanded uncertainty (A/A)} & \textbf{\boldmath$U$} 
  & \textbf{$3.1\times10^{-3}$} & \textbf{$1.8\times10^{-3}$} 
  & \textbf{$1.3\times10^{-3}$} & \textbf{$1.1\times10^{-3}$} 
  & \textbf{$9.0\times10^{-4}$} \\
\bottomrule
\end{tabular}
\end{table}

\end{document}